\documentclass[12pt]{article}
\usepackage{tikz}

\usetikzlibrary{decorations.pathmorphing}

\usepackage{jheppub}

\usepackage{pgfplots}
\usepackage{pgfplotstable}
\usepackage{pgfkeys}
\allowdisplaybreaks
\usepackage{ytableau}
\usepackage{comment}
\usepackage[vcentermath]{youngtab}
\usepackage{bm}
\usepackage{ulem}
\usepackage{xcolor}
\definecolor{nosaka}{rgb}{0.0, 0.7, 0.0}

\newcommand\be{\begin{equation}}
\newcommand\ee{\end{equation}}

\newcommand\Tr{\mathrm{Tr}}

\usepackage{braket}
\usepackage{subfigure}

\title{M2-M5 giant graviton expansions}
\abstract{
We examine the giant graviton expansions of the Coulomb and Higgs indices for the M2-brane SCFTs 
to find the closed-form expressions for the indices that encode the spectra of the $1/4$-BPS M5-brane giant gravitons and the $1/3$-BPS orbifold M5-brane giant gravitons. 
Consequently, we get exact forms of the twisted indices for the 6d $(2,0)$ theories describing a stack of $N$ M5-branes which generalize the unrefined indices. 
We confirm that they are also beautifully expanded with respect to the indices for the M2-brane giant gravitons 
which are obtained from the Coulomb and Higgs indices for the M2-brane SCFTs upon the change of variables. 
}
\author[a]{Hirotaka Hayashi,}
\emailAdd{h.hayashi@tokai.ac.jp}
\affiliation[a]{Department of Physics, School of Science, Tokai University,\\
4-1-1 Kitakaname, Hiratsuka-shi, Kanagawa 259-1292, Japan}
\author[b]{Tomoki Nosaka}
\emailAdd{nosaka@simis.cn}
\affiliation[b]{
Shanghai Institute for Mathematics and Interdisciplinary Sciences,\\
Block A, International Innovation Plaza, No.~657 Songhu Road, Yangpu District, Shanghai, China
}
\author[c]{and Tadashi Okazaki}
\emailAdd{tokazaki@seu.edu.cn}
\affiliation[c]{
Shing-Tung Yau Center of Southeast University,\\
Yifu Architecture Building, No.2 Sipailou, Xuanwu district, Nanjing, Jiangsu, 210096, China}

\begin{document}
\maketitle

\ytableausetup{boxsize=1.5mm}

\section{Introduction and summary}
\label{sec_intro}
Supersymmetric indices encode the spectra of the BPS operators in supersymmetric field theories 
and play a central role in understanding various open questions in theoretical physics and mathematics. 
In particular, for the low-energy effective world-volume theories on branes in string and M-theory, 
the indices can describe the spectra of quantum fluctuations of gravity theories according to the AdS/CFT correspondence \cite{Maldacena:1997re}.

The low-energy effective theories on the world-volume of a stack of $N$ coincident M2-branes (resp.~M5-branes) propagating in flat space 
are 3d $\mathcal{N}=8$ (resp.~6d $(2,0)$) superconformal field theories 
and they are the holographically dual descriptions of M-theory on $AdS_4\times S^7$ (resp.~$AdS_7\times S^4$). 
The supersymmetric indices for these theories in the large $N$ limit agree with 
the indices of the Kaluza-Klein (KK) modes of the massless fields in the supergravity background. 
The finite $N$ correction can be interpreted as the contributions from excitations of giant gravitons \cite{McGreevy:2000cw}, which carry large angular momenta and behave as a configuration of wrapped branes in the supergravity description. 
Recently it has been proposed that the supersymmetric indices can admit the ``giant graviton expansions'' \cite{Arai:2019xmp,Arai:2020qaj,Gaiotto:2021xce}, 
that is expansions of the ratios of the finite $N$ indices to the large $N$ indices with respect to the fugacities for the angular momenta carried by the giant gravitons. 
While several giant graviton expansions have been proposed, most of them are only checked by numerical methods. It is extremely desirable to have exact closed-forms for the indices to have more detailed study of the giant graviton expansions, including their rigorous proofs and the wall-crossing phenomena \cite{Gaiotto:2021xce,Lee:2022vig,Imamura:2022aua}.

In this paper, we examine the giant graviton expansions of the Coulomb indices and the Higgs indices for the theories of M2-branes probing the $A$-type singularities, 
including the ADHM theory \cite{deBoer:1996mp,deBoer:1996ck} and ABJ(M) theory \cite{Aharony:2008ug,Gaiotto:2008sd,Hosomichi:2008jb,Aharony:2008gk} 
as well as the Higgs indices for $\mathcal{N}=8$ and super Yang-Mills (SYM) theories. 
Remarkably, the finite $N$ corrections of the M2-brane indices can lead to understanding of the indices for mysterious M5-brane theories and vice versa.
By means of the grand canonical analysis for the Coulomb indices in \cite{Hayashi:2022ldo,Hayashi:2024jof}, 
we obtain the closed-form expressions for the $1/4$-BPS (resp.~$1/3$-BPS) indices of the M5-brane giant gravitons wrapping $5$-cycles as intersections of 
holomorphic surfaces with $S^7$ (resp.~$S^7/\mathbb{Z}_l$).\footnote{
Since the Coulomb index of the $U(N)$ ADHM theory with $l$ flavors is equivalent to that of the ${\cal N}=6$ $U(N)_l\times U(N)_{-l}$ ABJM theory \cite{Hayashi:2022ldo}, the giant graviton expansions can lead to the indices for the $1/4$-BPS M5-brane giant gravitons \cite{Herrero:2011bk,Lozano:2013ota} in the $AdS_4\times S^7/\mathbb{Z}_l$ that is holographically dual to the ABJM theory. 
}
Upon the change of variables they can be understood as certain special fugacity limits, which we call the \textit{twisted limit}, 
of the indices for 6d $(2,0)$ theories as supersymmetric partition functions on $S^1\times S^5$ (resp.~$S^1\times S^5/\mathbb{Z}_l$). 
Quite beautifully, the twisted indices also have the giant graviton expansions as inverse transformations 
in that the expansion coefficients are given by the indices of M2-brane giant gravitons which are obtained from the ADHM Coulomb indices upon the appropriate change of variables.
From the giant graviton expansions of the Higgs indices for the ADHM theory with $l$ flavors, we find the exact form of index for a single M5-brane giant graviton.
It is given by a product of the index for an M5-brane giant graviton with $l=1$ and the vacuum character of the $SU(l)_{1}$ WZW model of level $1$.
This is similar to the WZW model emergent from the M5-branes wrapped on Taub-NUT space \cite{Dijkgraaf:2007sw,Witten:2009at}.
Moreover, we also study the giant graviton expansions of the Higgs indices for $\mathcal{N}=8$ SYM theories 
which describe $N$ multiple D2-branes in Type IIA string theory. 
We find the closed-form expressions for the dual indices in terms of certain infinite products.  
It follows that they also admit the giant graviton expansions as inverse transformations. They provide us with prospective candidates for the indices of D4-brane giant gravitons. 

\subsection{Future works}

\begin{itemize}

\item While we demonstrate the giant graviton expansions for the Coulomb and Higgs indices for M2-brane SCFTs which are associated with the $1/4$-BPS M5-brane giant gravitons and the $1/3$-BPS orbifold M5-brane giant gravitons, it is desirable to figure out the giant graviton expansions for the $1/8$-BPS M5-brane giant gravitons \cite{Mikhailov:2000ya} and the $1/6$-BPS orbifold M5-brane giant gravitons \cite{Lozano:2013ota} by examining the giant graviton expansions for full supersymmetric indices. 

\item We are interested in studying the giant graviton expansions of supersymmetric indices for more general M2-brane SCFTs, 
including $\mathcal{N}=4$ supersymmetric $U(N)_k$ $\times$ $U(N)_0^{\otimes(p-1)}$ $\times$ $U(N)_{-k}$ quiver Chern-Simons matter theories 
describing a stack of $N$ M2-branes probing $(\mathbb{C}^2/\mathbb{Z}_{p}\times \mathbb{C}^2)/\mathbb{Z}_k$ \cite{Imamura:2008nn} 
whose large $N$ Higgs indices are calculated in \cite{Hayashi:2023txz}. 

\item It would be intriguing to study the giant graviton expansions of the line defect indices for the M2-brane SCFTs. 
Such giant graviton expansions for line defect indices in $\mathcal{N}=4$ SYM theory have been recently studied in \cite{Imamura:2024lkw,Beccaria:2024oif,Imamura:2024pgp,Beccaria:2024dxi,Hatsuda:2024uwt,Beccaria:2024lbt,Imamura:2024zvw}. 
We expect that the closed-form formulas of the Wilson line defect indices of the ADHM theories presented in \cite{Hayashi:2024jof} are crucial to proceed to further analysis. 

\item It would be interesting to directly derive our results on the gravity side. For the half-BPS D3-brane giant gravitons, the giant graviton expansions are addressed from the half-BPS bubbling geometry solutions \cite{Lin:2004nb} in Type IIB string theory by using the covariant quantization methods in \cite{Deddo:2024liu}. 

\end{itemize}

\subsection{Structure}
The paper is organized as follows. 
In section \ref{sec_giants} we review the M2- and M5-brane giant gravitons in M-theory. 
In section \ref{sec_index} we introduce the supersymmetric indices for the effective theories of M2-branes and M5-branes and their special fugacity limits. 
In section \ref{sec_gg_expansion} we study the giant graviton expansions of the indices for the M2-brane SCFTs. 
We present giant graviton expansions for the Coulomb indices and half-BPS indices for 3d $\mathcal{N}=4$ $U(N)$ ADHM theories  
and find the closed-form expressions for the indices for the $1/4$-BPS M5-brane giant gravitons in $S^7$ 
and the $1/3$-BPS orbifold M5-brane giant gravitons in $S^7/\mathbb{Z}_l$. 
Consequently, we find the exact forms of the twisted limits for the M5-brane indices. 
Also we find the giant graviton expansions of the Higgs indices for 3d $\mathcal{N}=8$ SYM theories as well as the index of single M5-brane giant graviton for 3d ${\cal N}=4$ $U(N)$ ADHM theory with $l$ flavors.
In appendix \ref{app_ggfromgc} we provide a heuristic argument to obtain 
the single sum giant graviton expansions of the Coulomb/Higgs indices of the M2-brane SCFTs analytically
from the grand canonical indices.
In appendix \ref{app_UNlHgg} we elaborate how to obtain the leading giant graviton coefficient in the Higgs indices of 3d ${\cal N}=4$ ADHM theory with $l$ flavors.


\section{M2-giants and M5-giants}
\label{sec_giants}
It was observed by McGreevy, Susskind and Toumbas \cite{McGreevy:2000cw} that 
gravitons in the near-horizon geometries produced by branes carrying large angular momenta behave as extended branes. 
They are called the giant gravitons. 
For $AdS_m\times S^n$ geometries the maximally supersymmetric giant gravitons are $(n-2)$-branes carrying angular momentum $P$ and wrapping $S^{n-2}$ in the $S^n$ 
in such a way that they are expanded into the spherical part of the background geometry as stable configurations. 
The stable configurations are characterized by a fixed size and an angular momentum $P$ 
and saturate the BPS bound $E=P/L$ for the energy $E$ where $L$ is the radius of the $S^n$. 
The angular momentum has a maximum value given by the quantized flux $N$ of the $n$-form field strength in the background geometry. 

\subsection{M5-brane giants}
\label{sec_M2giants}
When $(m,n)=(4,7)$, the 11d supergravity background geometry includes $AdS_4$ space, 
which is holographically dual to the M2-brane SCFT. 
For $AdS_4\times S^7$ geometry with $N$ units of $7$-form fluxes on $S^7$ 
there exist spherical giant gravitons as M5-branes wrapping $S^5$ $\subset$ $S^7$ \cite{McGreevy:2000cw}. 
As they preserve half of the supersymmetry, they are called the half-BPS M5-brane giant gravitons. 
More generally, one finds $1/4$- and $1/8$-BPS configurations of giant gravitons wrapping 
supersymmetric cycles in $S^7$ as intersections of the holomorphic surfaces with $S^7$ in $\mathbb{C}^4$ \cite{Mikhailov:2000ya}. 
We denote flat complex coordinates on $\mathbb{C}^4$ by $z_1$, $z_2$, $z_3$ and $z_4$ 
and describe $S^7$ as $|z_1|^2+|z_2|^2+|z_3|^2+|z_4|^2=1$. 
The holomorphic surface is described by the holomorphic equation of the form 
\begin{align}
\label{hol_giantM5}
f(z_1,z_2,z_3,z_4)&=
\sum_{a=1}^{4}c_a z_a=0, 
\end{align}
where $c_a$ are homogeneous coordinates on $\mathbb{CP}^3$. 
In other words, it is given by the zero locus of the holomorphic function. 
For the holomorphic equations depending only on two variables e.g. $f(z_1,z_3)=0$, one finds the configuration of the $1/4$-BPS M5-brane giant gravitons. 
For those with three or four variables, e.g. $f(z_1,z_2,z_3)=0$ or $f(z_1,z_2,z_3,z_4)=0$, we obtain the $1/8$-BPS M5-brane giant gravitons. 
The direction of motion of the giant gravitons is induced by the complex structure of $\mathbb{C}^4$. 

\subsection{M2-brane giants}
\label{sec_M5giants}
For $(m,n)=(7,4)$, the 
11d supergravity background contains $AdS_7$ factor, 
which is dual to 6d SCFT describing $N$ coincident M5-branes. 
The $AdS_7\times S^4$ geometry has a $4$-form flux of $N$ quanta on the $S^4$ 
and it involves M2-brane giant gravitons 
wrapping $S^2$ $\subset$ $S^4$. 
Such spherical configurations are the half-BPS M2-giant gravitons preserving half of the supersymmetry. 
Also there exist the $1/4$-BPS M2-giant gravitons associated with the holomorphic surface \cite{Mikhailov:2000ya}. 
We denote the coordinates on $\mathbb{C}^2$ by $Z_1$ and $Z_2$ 
and embed $S^4$ into $\mathbb{R}^5$ $=$ $\mathbb{C}^2\times \mathbb{R}$. 
When one considers three-dimensional cylinders of the form $\mathcal{C}\times \mathbb{R}$ in $\mathbb{C}^2\times \mathbb{R}$ 
where $\mathcal{C}$ is a generic holomorphic curve, the M2-giant gravitons wrapping the intersection of the 
cylinder with $S^4$ can preserve $1/4$ of the supersymmettry.

\subsection{Orbifold giants}
\label{sec_orbgiants}
Next consider the orbifold $AdS_4\times S^7/\mathbb{Z}_l$ geometry. 
The metric of $S^7$ can be written as a Hopf fibration $S^7$ $\hookleftarrow$ $\mathbb{CP}^3$ over $\mathbb{CP}^3$ 
\begin{align}
ds_{S^7}^2&=\left(d\tau-\sin^2\zeta d\chi+\cos^2\zeta\cos^2\frac{\theta_1}{2}d\varphi_1-\sin^2\zeta \cos^2\frac{\theta_2}{2}d\varphi_2\right)^2+ds_{\mathbb{CP}^3}^2,
\end{align}
where the fiber $0\le \tau\le 2\pi$ can be thought of as the overall phase of the complex coordinates $z_a$ and 
\begin{align}
ds_{\mathbb{CP}^3}^2
&=d\zeta^2+\cos^2\zeta\sin^2\zeta \left(d\chi+\cos^2\frac{\theta_1}{2}d\varphi_1+\cos^2\frac{\theta_2}{2}d\varphi_2 \right)^2
\nonumber\\
&+\frac14 \cos^2\zeta \left(d\theta_1^2+\sin^2\theta_1 d\varphi_1^2 \right)+\frac14\sin^2\zeta \left(d\theta_2^2+\sin^2\theta_2 d\varphi_2^2 \right)
\end{align}
is the Fubini-Study metric of $\mathbb{CP}^3$ with 
radial coordinates $(\zeta,\theta_1,\theta_2)$ and angular coordinates $(\chi,\varphi_1,\varphi_2)$. 

The orbifold $S^7/\mathbb{Z}_l$ can be constructed as the orbifold identification on the Hopf fibre with $\tau$ $\sim$ $\tau+\frac{2\pi}{l}$. 
It is argued in \cite{Lozano:2013ota} 
that it can support M5-brane giant gravitons wrapping the intersections of the holomorphic surfaces in $\mathbb{C}^4/\mathbb{Z}_l$ with the orbifold space $S^7/\mathbb{Z}_l$ 
(see also \cite{Aharony:2008ug,Herrero:2011bk}). 
They are called the ``orbifold giant gravitons''. 
Since $\tilde{\tau}$ $:=$ $l \tau$ has the usual $2\pi$ periodicity, it is identified with an effective orbifold fibre. 
This is compatible with the fact that the M5-brane giant graviton carries the mod $l$ wrapping number and the $7$-form flux in the presence of the $\mathbb{Z}_l$ orbifold action is $l$ times as the ordinary flux. 
As we will see in section \ref{sec_gg_expansion}, the supersymmetric indices for M2-branes probing the $\mathbb{Z}_l$ orbifold has expansions 
with respect to giant gravitons of the mod $l$ wrapping numbers 
so that the expansion coefficients can be identified with the supersymmetric indices for the orbifold M5-brane giant gravitons. 

On the world-volume of the orbifold giant gravitons the multi-valued Killing spinors are supported and some of the supersymmetry of M5-brane giant gravisons are killed.
For generic holomorphic surface of the form (\ref{hol_giantM5}) they preserve $1/6$ of the supersymmetry, $4$ supercharges. 
However, when the holomorphic equations only depend on certain two variables, i.e. $f(z_1,z_3)=0$, they have enhanced $1/3$ of the supersymmetry\footnote{
Note that there also exist $1/6$-BPS orbifold M5-brane giant gravitons associated with the holomorphic equations with two variables \cite{Lozano:2013ota}.
}
and when they only involve a single variable, the M5-brane giant graviton becomes the $1/2$-BPS configuration \cite{Lozano:2013ota}. 
These giant gravitons should capture the finite $N$ corrections of the supersymmetric indices for the M2-brane SCFTs. 
In fact, we will find the BPS indices for the orbifold M5-brane giant gravitons with enhanced supersymmetry 
as the coefficients in the giant graviton expansions of the supersymmetic indices for the M2-brane SCFTs. 

\subsection{D4-brane giants}
\label{sec_d4giants}
Upon the compactification of M-theory geometry $AdS_4\times S^7/\mathbb{Z}_k$ 
on the fibre of the Hopf fibration $S^7$ $\hookleftarrow$ $\mathbb{CP}^3$ over $\mathbb{CP}^3$, 
one obtains Type IIA string theory on $AdS_4\times \mathbb{CP}^3$ \cite{Nilsson:1984bj}. 
The orbifold M5-brane giant gravitons can result in D4-branes\footnote{
Although the orbifold M5-brane giant gravitons may become motionless NS5-branes with non-trivial D0-brane charge, 
they will have no interpretation as giant gravitons.}
carrying D0-brane charge and angular momentum in the projective space. 
Although there still remain many open questions about the detailed properties of these objects, several interesting examples are investigated in the literature \cite{Aharony:2008ug,Aharony:2008gk,Berenstein:2008dc,Berenstein:2009sa,Giovannoni:2011pn,Hirano:2012vz,Herrero:2011bk,Lozano:2013sra,Lozano:2013ota}. 
For example, the $1/2$-BPS maximal M5-brane giant gravitons become D4-branes wrapping $\mathbb{CP}^2$ $\subset$ $\mathbb{CP}^3$ 
and their dual operators are identified with the dibaryon operators \cite{Aharony:2008ug,Berenstein:2008dc,Berenstein:2009sa}. 
In the following, we will also find the giant graviton expansions of the Higgs indices for $\mathcal{N}=8$ SYM theories, 
whose expansion coefficients are expected to be interpreted as the indices for the D4-brane giant gravitons. 

\section{Supersymmetric indices}
\label{sec_index}
We introduce conventions and definitions for supersymmetric indices for M2-brane SCFTs and M5-brane SCFTs. 
New features in our discussion include the ``twisted limit'', 
that is a special fugacity limit of the supersymmetric indices of 6d $(2,0)$ theories  
which generalizes the unrefined limit \cite{Kim:2012ava}. 

\subsection{M2-brane indices}
The low-energy effective theory of a stack of $N$ coincident M2-branes probing $\mathbb{C}^4$ is a 3d $\mathcal{N}=8$ maximally superconformal field theory. 
The 3d $\mathcal{N}=8$ superconformal group is $OSp(8|4)$ 
whose bosonic subgroup is $USp(4)$ $\times$ $SO(8)$ $\cong$ $SO(2,3)$ $\times$ $SO(8)$. 
There is a rotation generator $j_{12}$ of $SO(3)$ $\subset$ $SO(2,3)$, 
the Hamiltonian $h$ of $SO(2)$ $\subset$ $SO(2,3)$ 
and four Cartan generators $r_{12}$, $r_{34}$, $r_{56}$, $r_{78}$ of the $SO(8)$ R-symmetry. 

The supersymmetric index of the M2-brane SCFT can be defined by \cite{Bhattacharya:2008zy}
\begin{align}
\label{def_M2index}
&
\mathcal{I}^{\textrm{M2 SCFT}}[\mathbb{C}^4;N](t,z,x;q)
\nonumber\\
&:=\Tr (-1)^F 
q^{j_{12}+\frac14(r_{12}+r_{34}+r_{56}+r_{78})}
t^{-r_{12}+r_{34}-r_{56}+r_{78}}
z^{r_{12}-r_{56}}
x^{r_{34}-r_{78}}. 
\end{align}
The states which contribute to the index saturate the BPS bound
\begin{align}
\left\{ \mathcal{Q},\mathcal{Q}^{\dag}\right\}
&\sim
h-j_{12}-\frac12(r_{12}+r_{34}+r_{56}+r_{78})=0, 
\end{align}
where the charges of the chosen supercharge $\mathcal{Q}$ are given by
\begin{align}
\begin{array}{c|cccccc}
&h&j_{12}&r_{12}&r_{34}&r_{56}&r_{78} \\ \hline
\mathcal{Q}&\frac12&-\frac12&\frac12&\frac12&\frac12&\frac12 \\
\end{array}.
\end{align}

The world-volume theories of a stack of $N$ M2-branes probing the quotient singularity $\mathbb{C}^2/\Gamma$, where $\Gamma$ is a discrete subgroup of $SU(2)$, preserve $\mathcal{N}=4$ supersymmetry and $SU(2)_H\times SU(2)_C$ $\subset$ $SU(4)$ R-symmetry. 
We define the $\mathcal{N}=4$ supersymmetric index by\footnote{
See \cite{Okazaki:2019ony,Hayashi:2022ldo} for the definition and convention.
}
\begin{align}
\label{def_N=4index}
\mathcal{I}^{\textrm{M2 SCFT}}[\mathbb{C}^2\times \mathbb{C}^2/\Gamma;N](t,z,x;q)
&:=\Tr (-1)^F 
q^{j_{12}+\frac14(H+C)}
t^{H-C}
z^{f_C}
x^{f_H}. 
\end{align}
Here $H$ and $C$ are the Cartan generators of the $SU(2)_H$ and $SU(2)_C$. 
$f_C$ and $f_H$ are the Cartan generators of the topological symmetry $G_{C}$ and the flavor symmetry $G_{H}$. 

Note that the index (\ref{def_M2index}) can be rewritten as the $\mathcal{N}=4$ supersymmetric index with 
\begin{align}
C&:=r_{12}+r_{56},&  
H&:=r_{34}+r_{78}, \nonumber\\
f_C&:=r_{12}-r_{56},& 
f_H&:=r_{34}-r_{78}.
\end{align}
In fact, one can get the index (\ref{def_M2index}) from $\mathcal{N}=4$ ADHM theory of gauge group $U(N)$ with one flavor. 

\subsubsection{Coulomb and Higgs limits}
$\mathcal{N}\ge4$ supersymmetric field theories contain two types of half-BPS local operators. 
One is the Coulomb branch operator consisting of adjoint scalar fields in the vector multiplet and the monopole operators. 
The other is the Higgs branch operator as gauge invariant polynomial in the scalar fields in the hypermultiplet. 

The indices for the spectra of the Coulomb and Higgs branch operators 
can be obtained by taking the following fugacity limits of $\mathcal{N}=4$ supersymmetric indices \cite{Razamat:2014pta}: 
\begin{align}
\label{def_Cindex}
\mathcal{I}^{\textrm{M2 SCFT}}_{C}(\mathfrak{t},z)&:=
\lim_{
\begin{smallmatrix}
\mathfrak{t}=q^{\frac14}t^{-1}: \textrm{fixed}, \\
q\rightarrow 0, t\rightarrow 0, \\
\end{smallmatrix}
}
\mathcal{I}^{\textrm{M2 SCFT}}(t,z,x;q), \\
\label{def_Hindex}
\mathcal{I}^{\textrm{M2 SCFT}}_{H}(\mathfrak{t},x)&:=
\lim_{
\begin{smallmatrix}
\mathfrak{t}=q^{\frac14}t: \textrm{fixed}, \\
q\rightarrow 0, t\rightarrow \infty, \\
\end{smallmatrix}
}
\mathcal{I}^{\textrm{M2 SCFT}}(t,z,x;q). 
\end{align}
The Coulomb index (\ref{def_Cindex}) (resp.~Higgs index (\ref{def_Hindex})) depends on a pair of fugacities $\mathfrak{t}$ and $z$ (resp.~$\mathfrak{t}$ and $x$). 
In the unflavored limit $z\rightarrow 1$ (resp.~$x\rightarrow 1$), 
the Coulomb (resp.~Higgs) indices become the unrefined Hilbert series for the chiral rings of the holomorphic functions on the Coulomb (resp.~Higgs) branches 
\cite{Cremonesi:2013lqa,Cremonesi:2014kwa,Cremonesi:2014vla}. 
Instead of these fugacities, we also introduce 
\begin{align}
x_1&:=\mathfrak{t}z^{-1},\qquad x_2:=\mathfrak{t}z
\end{align}
for the Coulomb indices and
\begin{align}
x_1&:=\mathfrak{t}x^{-1},\qquad x_2:=\mathfrak{t}x
\end{align}
for the Higgs indices. 

The unflavored limit is obtained by taking
\begin{align}
\label{unflavor_lim}
x_2\rightarrow x_1. 
\end{align}

Furthermore, when $x_2\rightarrow 0$, the Coulomb or Higgs index reduces to the half-BPS index \cite{Bhattacharyya:2007sa}
\begin{align}
\label{1/2BPSM2_lim}
\mathcal{I}^{\textrm{M2 SCFT}}_{\textrm{$\frac12$BPS},C/H}(x_1)
&:=\lim_{x_2\rightarrow 0}\mathcal{I}^{\textrm{M2 SCFT}}_{C/H}(x_1;x_2). 
\end{align}
While the half-BPS limits obtained from the Coulomb and Higgs indices are generically different, 
they give rise to the same indices for self-mirror theory. 

\subsubsection{$N=1$}
There exist several dual UV descriptions which describe a single M2-brane moving in $\mathbb{C}^4$. 
For example, it can be described by a free matter theory consisting of hypermultiplet and twisted hypermultiplet. 
The index is given by \cite{Bhattacharya:2008zy}
\begin{align}
\label{index_1M2}
\mathcal{I}^{\textrm{free matter}}[\mathbb{C}^4;N=1](t,z,x;q)
&=\mathrm{PE}[ i^{\textrm{free matter}}[\mathbb{C}^4;N=1](t,z,x;q)], 
\end{align}
where 
\begin{align}
i^{\textrm{free matter}}[\mathbb{C}^4;N=1](t,z,x;q)
&=\frac{q^{\frac14}t(1-q^{\frac12}t^{-2})(x+x^{-1})+q^{\frac14}t^{-1}(1-q^{\frac12}t^{2})(z+z^{-1})}
{1-q}
\end{align}
and $\mathrm{PE}$ is the plethystic exponential 
\begin{align}
\mathrm{PE}[f(x_i)]&=\exp
\left(
\sum_{n=1}^{\infty}\frac{1}{n} f(x_i^n)
\right). 
\end{align}
The index (\ref{index_1M2}) can be written as the infinite product
\begin{align}
\label{index_1M2_prod}
\frac{(q^{\frac34}tz;q)_{\infty} (q^{\frac34}tz^{-1};q)_{\infty}(q^{\frac34}t^{-1}x;q)_{\infty} (q^{\frac34}t^{-1}x^{-1};q)_{\infty}}
{(q^{\frac14}t^{-1}z;q)_{\infty} (q^{\frac14}t^{-1}z^{-1};q)_{\infty}(q^{\frac14}tx;q)_{\infty} (q^{\frac34}tx^{-1};q)_{\infty}}, 
\end{align}
where 
\begin{align}
(x;q)_{n}=\prod_{i=0}^{n-1}(1-xq^i)
\end{align}
is the $q$-Pochhammer symbol.

Since the theory is self-mirror, the Coulomb index is equivalent to the Higgs index. 
One finds
\begin{align}
\label{Cindex_1M2}
\mathcal{I}^{\textrm{free matter}}_{C}[\mathbb{C}^4;N=1](x_1,x_2)&=
\mathcal{I}^{\textrm{free matter}}_{H}[\mathbb{C}^4;N=1](x_1,x_2)
\nonumber\\
&=\frac{1}{(1-x_1)(1-x_2)}. 
\end{align}

The half-BPS index is simply given by 
\begin{align}
\mathcal{I}^{\textrm{free matter}}_{\textrm{$\frac12$BPS}}[\mathbb{C}^4;N=1](x_1)
&=\frac{1}{1-x_1}. 
\end{align}
This agrees with the result in \cite{Sheikh-Jabbari:2009vjj}. 

\subsubsection{Large $N$}
In the large $N$ limit 
the index for $N$ M2-branes propagating in $\mathbb{C}^4$ encodes the spectra for the KK modes of supergravitons on the supergravity background $AdS_4\times S^7$. 
The spectrum of the supergravitons was addressed in \cite{Casher:1984ym,Gunaydin:1985tc,Aharony:1998rm,Minwalla:1998rp}. 
The supergravitons belong to the supershort multiplets of the superconformal algebra $\mathfrak{osp}(8|4)$. 
It is given by \cite{Bhattacharya:2008zy}
\begin{align}
\label{index_largeNM2}
\mathcal{I}^{\textrm{$AdS_4\times S^7$ KK}}[\mathbb{C}^4;N=\infty](t,z,x;q)
&=\mathrm{PE}[ i^{\textrm{$AdS_4\times S^7$ KK}}[\mathbb{C}^4;N=\infty](t,z,x;q) ], 
\end{align}
where 
\begin{align}
&
i^{\textrm{$AdS_4\times S^7$ KK}}[\mathbb{C}^4;N=\infty](t,z,x;q)
\nonumber\\
&=\frac{(1-q^{\frac34}tz) (1-q^{\frac34}tz^{-1})(1-q^{\frac34}t^{-1}x) (1-q^{\frac34}t^{-1}x^{-1})}
{(1-q^{\frac14}t^{-1}z) (1-q^{\frac14}t^{-1}z^{-1})(1-q^{\frac14}tx) (1-q^{\frac14}tx^{-1})(1-q^2)}
-\frac{1-q+q^2}{(1-q)^2}
\end{align}
is the single particle index. 

In the Coulomb limit or equivalently Higgs limit the large $N$ index becomes 
\begin{align}
\label{Cindex_largeNM2}
\mathcal{I}^{\textrm{$AdS_4\times S^7$ KK}}_{C}[\mathbb{C}^4;N=\infty](x_1;x_2)
&=\mathcal{I}^{\textrm{$AdS_4\times S^7$ KK}}_{H}[\mathbb{C}^4;N=\infty](x_1;x_2)
\nonumber\\
&=\prod_{n=1}^{\infty}\frac{1}{1-x_1^nx_2^n}
\prod_{m=0}^{\infty}\prod_{n=1}^{\infty}
\frac{1}{(1-x_1^{m+n}x_2^{m})(1-x_1^{m}x_2^{m+n})}. 
\end{align}

The half-BPS index is evaluated as
\begin{align}
\mathcal{I}^{\textrm{$AdS_4\times S^7$ KK}}_{\textrm{$\frac12$BPS}}[\mathbb{C}^4;N=\infty](x_1)
&=\prod_{n=1}^{\infty}\frac{1}{1-x_1^n}. 
\end{align}
Again this reproduces the result in \cite{Sheikh-Jabbari:2009vjj}. 

\subsection{M5-brane indices}
The low-energy effective theory of a stack of coincident M5-branes probing $\mathbb{R}^5$ is a 6d $(2,0)$ maximally superconformal field theories. 
The superconformal group of 6d $(2,0)$ theory is $OSp(8^*|4)$ 
whose bosonic subgroup is $SO^*(8)$ $\times$ $USp(4)$ $\cong$ $SO(2,6)$ $\times$ $SO(5)$. 
There are three rotation generators $J_{12}$, $J_{34}$, $J_{56}$ of $SO(6)$ $\subset$ $SO(2,6)$, 
the Hamiltonian $H$ of $SO(2)$ $\subset$ $SO(2,6)$ 
and two Cartan generators $R_{12}$, $R_{34}$ of the $SO(5)$ R-symmetry. 

We define the supersymmetric index of M5-brane SCFT by \cite{Bhattacharya:2008zy}
\begin{align}
\label{def_M5index}
&\mathcal{I}^{\textrm{M5 SCFT}}[\mathbb{R}^5;N](u,y_1,y_2,y_3;p)
\nonumber\\
&:=\Tr (-1)^F 
p^{\frac32(R_{12}+R_{34})+J_{12}+J_{34}+J_{56}}
u^{R_{12}-R_{34}}
y_1^{J_{12}}y_2^{J_{34}}y_3^{J_{56}}, 
\end{align}
where $y_1y_2y_3=1$. 
The states that contribute to the index obey the BPS bound
\begin{align}
\left\{Q,Q^{\dag}\right\}&
\sim
H-(J_{12}+J_{34}+J_{56})-2(R_{12}+R_{34})=0, 
\end{align}
where the chosen supercharge $Q$ carries the following charges
\begin{align}
\begin{array}{c|cccccc}
&H&J_{12}&J_{34}&J_{56}&R_{12}&R_{34} \\ \hline
Q&\frac12&-\frac12&-\frac12&-\frac12&\frac12&\frac12 \\
\end{array}. 
\end{align}

The index can be viewed as a supersymmetric partition function on $S^1\times S^5$ with a twisted periodicity.
Motivated from the ideas of \cite{Douglas:2010iu,Lambert:2010iw}, the index was computed in \cite{Lockhart:2012vp,Kim:2012qf,Kim:2012ava,Kim:2013nva} from the partition function of 5d $\mathcal{N}=1^*$ theory on a squashed $S^5$, where the fugacity $u$ corresponds to a mass parameter $m$ for the adjoint field while the fugacities $y_1$, $y_2$ and $y_3$ describe squashing parameters coupled to the Cartan generators of the isometry group of $S^5$

Furthermore, the general index was proposed as a supersymmetric partition function on $S^1\times S^5/\mathbb{Z}_l$ \cite{Kim:2012tr,Kim:2013nva}. 

\subsubsection{Twisted limit}
There exist an interesting limit of the general M5-brane index (\ref{def_M5index}) that plays a significant role in the giant graviton expansions 
of the Coulomb/Higgs indices of the M2-brane SCFTs. 
We introduce 
\begin{align}
\mathcal{I}^{\textrm{M5 SCFT}}_{\textrm{twist}}[\mathbb{R}^5;N](x_1;x_2)
&:=
\lim_{
\begin{smallmatrix}
x_1=p^{\frac32}u: \textrm{fixed}, \\
x_2=p^{-\frac12}u^{-1}y_2: \textrm{fixed}, \\
p,y_1, y_3\rightarrow 0, u, y_2\rightarrow \infty \\
\end{smallmatrix}
}
\mathcal{I}^{\textrm{M5 SCFT}}[\mathbb{R}^5;N](u,y_1,y_2,y_3;p), 
\label{IM5twisted}
\end{align}
which we call the \textit{twisted limit}. 
In the twisted limit we fix ratios $p^{\frac32}/u^{-1}$ and $p/y_2^{-1}$ by first introducing $x_1$ $=$ $p^{\frac32}u$ and $x_2$ $=$ $p^{-\frac12}u^{-1}y_2$ and keep them finite 
while taking $q,y_1,y_3$ $\rightarrow 0$ and $u,y_2$ $\rightarrow \infty$. 
One way to realize this limit is to 
specialize the fugacities
$(u,y_1,y_2,y_3;p)$ as
\begin{align}
{\cal I}^{\text{M5 SCFT}}[\mathbb{R}^5;N](
t^{-2}x_1^{\frac{1}{2}}x_2^{-1},
t^{\frac{2}{3}}x_1^{-\frac{1}{3}}x_2^{-\frac{1}{6}},
t^{-\frac{4}{3}}x_1^{\frac{2}{3}}x_2^{\frac{1}{3}},
t^{\frac{2}{3}}x_1^{-\frac{1}{3}}x_2^{-\frac{1}{6}}
;
t^{\frac{4}{3}}x_1^{\frac{1}{3}}x_2^{\frac{2}{3}}
),
\label{twistedlimitbyt}
\end{align}
and take the limit $t\rightarrow 0$ while $x_1,x_2$ kept finite.
As we will see in section \ref{sec_gg_expansion} (see \eqref{M2M5dictionaryforz56}), 
this parameterization 
arises
naturally in the holographic interpretation of the giant graviton expansion of the Coulomb limit of the 
M2-brane indices \cite{Arai:2020uwd}. 

The resulting twisted index depends on a pair of fugacities $x_1$ and $x_2$ 
so that it gives more information than the unrefined limit \cite{Kim:2012ava}. 
The unrefined limit is simply obtained by specializing $x_2$ to $1$ or equivalently setting $y_2$ to $p^{\frac12}u$. 
Indeed, according to the original definition \eqref{def_M5index} with trace over the states, the M5-brane index 
specialized as \eqref{twistedlimitbyt} is written as
\begin{align}
&{\cal I}^{\text{M5 SCFT}}[\mathbb{R}^5;N](
t^{-2}x_1^{\frac{1}{2}}x_2^{-1},
t^{\frac{2}{3}}x_1^{-\frac{1}{3}}x_2^{-\frac{1}{6}},
t^{-\frac{4}{3}}x_1^{\frac{2}{3}}x_2^{\frac{1}{3}},
t^{\frac{2}{3}}x_1^{-\frac{1}{3}}x_2^{-\frac{1}{6}}
;
t^{\frac{4}{3}}x_1^{\frac{1}{3}}x_2^{\frac{2}{3}}
)\nonumber \\
&=
\text{Tr}(-1)^Ft^{2\{Q',(Q')^\dagger\}-2\{Q,Q^\dagger\}}x_1^{R_{12}+J_{34}}x_2^{2R_{34}+\frac{J_{12}}{2}+J_{34}+\frac{J_{56}}{2}},
\end{align}
where $Q'$ is another supercharge carrying the following charges
\begin{align}
\begin{array}{c|cccccc|c|c|}
&H&J_{12}&J_{34}&J_{56}&R_{12}&R_{34}&R_{12}+J_{34}&2R_{34}+\frac{J_{12}}{2}+J_{34}+\frac{J_{56}}{2} \\ \hline
Q'&\frac12&\frac12&-\frac12&\frac12&\frac12&-\frac12&0&-1 \\
\end{array}, 
\end{align}
which satisfies
\begin{align}
\{Q',(Q')^\dagger\}
\sim
H+J_{12}-J_{34}+J_{56}-2R_{12}+2R_{34}.
\end{align}
Since $R_{12}+J_{34}$ commutes with both $Q$ and $Q'$, the M5-brane index \eqref{twistedlimitbyt} with $x_2$ set to $1$ is actually independent of the parameter $t$ controlling the twisted limit.
Namely, for $x_2=1$, the M5-brane index in the twisted limit \eqref{IM5twisted} is the same as the M5-brane index before taking the limit \eqref{twistedlimitbyt} for $x_2=1$ and any choice of $t$.
In particular, by setting $t=x_1^{\frac{1}{2}}$ we obtain
\begin{align}
\label{IM5unref}
\mathcal{I}^{\textrm{M5 SCFT}}_{\textrm{unref}}[\mathbb{R}^5;N](x_1)
&=\mathcal{I}^{\textrm{M5 SCFT}}_{\textrm{twist}}[\mathbb{R}^5;N](x_1,1)
\nonumber\\
&={\cal I}^{\text{M5 SCFT}}[\mathbb{R}^5;N](x_1^{-\frac{1}{2}},1,1,1;x_1), 
\end{align}
which reproduces the unrefined limit proposed in \cite{Kim:2012ava}.

Also, 
when we only keep $x_1$ finite and take $p\rightarrow 0$, $u\rightarrow \infty$ of the M5-brane index, we find the half-BPS index \cite{Bhattacharya:2008zy}
\begin{align}
\mathcal{I}^{\textrm{M5 SCFT}}_{\textrm{$\frac12$BPS}}[\mathbb{R}^5;N](x_1)&:=
\lim_{
\begin{smallmatrix}
x_1=p^{\frac32}u: \textrm{fixed} \\
p\rightarrow0, u\rightarrow\infty \\
\end{smallmatrix}
}
\mathcal{I}^{\textrm{M5 SCFT}}[\mathbb{R}^5;N](u,y_1,y_2,y_3;p). 
\end{align}
The result turns out to be independent of the values of $y_1,y_2,y_3$. 
For the maximally supersymmetric configuration of M5-branes, 
it counts the BPS states with $16$ supercharges. 

\subsubsection{$N=1$}
For $N=1$ we have the world-volume theory on a single M5-brane as the 6d $(2,0)$ theory of a free tensor multiplet. 
The index reads \cite{Bhattacharya:2008zy}
\begin{align}
\label{index_1M5}
\mathcal{I}^{\textrm{free tensor}}[\mathbb{R}^5;N=1](u,y_1,y_2,y_3;p)
&=\mathrm{PE}[ i^{\textrm{free tensor}}[\mathbb{R}^5;N=1](u,y_1,y_2,y_3;p)], 
\end{align}
where 
\begin{align}
\label{sindex_1M5}
i^{\textrm{free tensor}}[\mathbb{R}^5;N=1](u,y_1,y_2,y_3;p)
&=\frac{p^{\frac32}(u+u^{-1})-p^2(y_1^{-1}+y_2^{-1}+y_3^{-1})+p^3}
{(1-py_1)(1-py_2)(1-py_3)}. 
\end{align}

In the twisted limit the single particle index (\ref{sindex_1M5}) reduces to
\begin{align}
\label{sTindex_1M5}
i^{\textrm{free tensor}}_{\textrm{twist}}[\mathbb{R}^5;N=1](x_1;x_2)
&=\frac{x_1}{1-x_1 x_2}. 
\end{align}
Hence we get
\begin{align}
\label{Tindex_1M5}
\mathcal{I}^{\textrm{free tensor}}_{\textrm{twist}}[\mathbb{R}^5;N=1](x_1;x_2)
&=\prod_{n=0}^{\infty}
\frac{1}{1-x_1^{n+1}x_2^n}. 
\end{align}

In the unrefined limit one finds \cite{Kim:2012ava}
\begin{align}
\mathcal{I}^{\textrm{free tensor}}_{\textrm{unref}}[\mathbb{R}^5;N=1](p)
&=\prod_{n=1}^{\infty}\frac{1}{1-p^{n}}. 
\end{align}

In the half-BPS limit the single particle index (\ref{sindex_1M5}) simply becomes $x_1$. 
Hence the half-BPS index is given by \cite{Bhattacharyya:2007sa}
\begin{align}
\mathcal{I}^{\textrm{free tensor}}_{\textrm{$\frac12$BPS}}[\mathbb{R}^5;N=\infty](x_1)&=\frac{1}{1-x_1}. 
\end{align}

In \cite{Kim:2013nva} the general index for the 6d $(2,0)$ free tensor multiplet theory on $S^1\times S^5/\mathbb{Z}_l$ was addressed 
from the partition function on lens space $S^5/\mathbb{Z}_l$ via the supersymmetric localization. 
It takes the form 
\begin{align}
\label{gindex_1M5}
\mathcal{I}^{\textrm{free tensor $\mathbb{Z}_l$}}[\mathbb{R}^5;N=1](u,y_1,y_2,y_3;p)
&=\mathrm{PE}[ i^{\textrm{free tensor $\mathbb{Z}_l$}}[\mathbb{R}^5;N=1](u,y_1,y_2,y_3;p) ], 
\end{align}
where 
\begin{align}
\label{sgindex_1M5}
i^{\textrm{free tensor $\mathbb{Z}_l$}}[\mathbb{R}^5;N=1](u,y_1,y_2,y_3;p)
&=i_{\textrm{inst}}(u,y_1,y_2,y_3;p)\frac{p^{l}y_1^l}{1-p^l y_1^l}
\nonumber\\
&+(y_1,y_2,y_3\rightarrow y_2,y_3,y_1)
+(y_1,y_2,y_3\rightarrow y_3,y_1,y_2)
\end{align}
and 
\begin{align}
i_{\textrm{inst}}(u,y_1,y_2,y_3;p)
&=p^{-\frac12}u^{-1}y_1^{-2}
\frac{(1-\frac{p^{\frac12}u}{y_2}) (1-\frac{p^{\frac12}u}{y_3})}
{(1-\frac{y_2}{y_1})(1-\frac{y_3}{y_1})}
\end{align}
is the single particle index for a half-BPS instanton on $\mathbb{R}^{4,1}$ \cite{Kim:2011mv}. 

In the twisted limit, the single particle index (\ref{sgindex_1M5}) becomes
\begin{align}
\label{sgTindex_1M5}
i^{\textrm{free tensor $\mathbb{Z}_l$}}_{\textrm{twist}}[\mathbb{R}^5;N=1](x_1;x_2)
&=\frac{x_1^l x_2^{l-1}}{1-x_1^l x_2^l}. 
\end{align}
Note that the twisted index is simply obtained from (\ref{sTindex_1M5}) by performing the $\mathbb{Z}_l$ projection \cite{Gustavsson:2018sgi}
\begin{align}
i^{\textrm{free tensor $\mathbb{Z}_l$}}_{\textrm{twist}}[\mathbb{R}^5;N=1](x_1;x_2)
&=\frac{1}{l}\sum_{m=0}^{l-1}
i^{\textrm{free tensor}}_{\textrm{twist}}[\mathbb{R}^5;N=1](e^{\frac{2\pi i m}{l}}x_1;e^{\frac{2\pi i m}{l}}x_2). 
\end{align}
However, we expect that the general index has more structures. 
Since the lens space has non-trivial homotopy group $\pi_1(S^5/\mathbb{Z}_l)$ $=$ $\mathbb{Z}_l$ that results in $l$ degenerate vacua, 
the total partition function on lens space $S^5/\mathbb{Z}_l$ will be given by a sum of $l$ separate sectors 
where each sector is distinguished by a certain phase factor. 
From the giant graviton expansion we will see the expected form of the general index in the twisted limit. 

\subsubsection{Large $N$}
The large $N$ index for the 6d $(2,0)$ $A_{N-1}$ theory describing $N$ M5-branes in flat space 
is identified with the index for the KK modes on the supergravity background $AdS_7\times S^4$. 
The supergravitons fit into supermultiplets of the superconformal algebra $\mathfrak{osp}(8^*|4)$. 
The spectrum was investigated in \cite{Gunaydin:1984wc,Aharony:1998rm,Minwalla:1998rp}.
It is given by \cite{Bhattacharya:2008zy}
\begin{align}
\label{index_largeNM5}
\mathcal{I}^{\textrm{$AdS_7\times S^4$ KK}}[\mathbb{R}^5;N=\infty](u,y_1,y_2,y_3;p)
&=\mathrm{PE}[ i^{\textrm{$AdS_7\times S^4$ KK}}[\mathbb{R}^5;N=\infty](u,y_1,y_2,y_3;p)], 
\end{align}
where 
\begin{align}
\label{sindex_largeNM5}
&
i^{\textrm{$AdS_7\times S^4$ KK}}[\mathbb{R}^5;N=\infty](u,y_1,y_2,y_3;p)
\nonumber\\
&=\frac{p^{\frac32}(u+u^{-1})-p^2(y_1^{-1}+y_2^{-1}+y_3^{-1})+p^4(y_1+y_2+y_3)-p^{\frac92}(u+u^{-1})}
{(1-up^{\frac32})(1-u^{-1}p^{\frac32})(1-py_1)(1-py_2)(1-py_3)}. 
\end{align}

In the twisted limit, we obtain the single particle index  
\begin{align}
i^{\textrm{$AdS_7\times S^4$ KK}}_{\textrm{twist}}[\mathbb{R}^5;N=\infty](x_1;x_2)
&=\frac{x_1}{(1-x_1)(1-x_1 x_2)}
\end{align}
and the multiparticle gravity index 
\begin{align}
\label{Tindex_largeNM5}
\mathcal{I}^{\textrm{$AdS_7\times S^4$ KK}}_{\textrm{twist}}[\mathbb{R}^5;N=\infty](x_1;x_2)
&=\prod_{k=1}^{\infty}\prod_{n=0}^{\infty}
\frac{1}{1-x^{n+k}x_2^n}. 
\end{align}

The unrefined large $N$ index is given by the MacMahon function 
or equivalently the generating function for plane partitions
\begin{align}
\mathcal{I}^{\textrm{$AdS_7\times S^4$ KK}}_{\textrm{unref}}[\mathbb{R}^5;N=\infty](p)
&=\prod_{n=1}^{\infty}\frac{1}{(1-p^n)^{n}}. 
\end{align}

The large $N$ half-BPS index is given by \cite{Bhattacharyya:2007sa}
\begin{align}
\mathcal{I}^{\textrm{$AdS_7\times S^4$ KK}}_{\textrm{$\frac12$BPS}}[\mathbb{R}^5;N=\infty](x_1)&=
\prod_{n=1}^{\infty}\frac{1}{1-x_1^n}. 
\end{align}

\section{Giant graviton expansions}
\label{sec_gg_expansion}
While the large $N$ limits of the supersymmetric indices of the gauge theoreis for a stack of $N$ coincident branes in string/M-theory 
exactly coincide with the gravity indices for the Kaluza-Klein (KK) excitations of the supergravitons on the supergravity background, 
for finite $N$ the gravity indices should contain the contributions from the giant gravitons whose angular momenta are of order $N$. 
The giant graviton expansions are 
the expansions of the ratios of the finite $N$ supersymmetric indices to the large $N$ indices with respect to the fugacities coupled to the angular momenta carried by the giant gravitons \cite{Arai:2019xmp,Arai:2020qaj,Gaiotto:2021xce}. 
Such expansions are highly non-trivial as most of them are merely checked by the numerical methods except for the case with the exact forms. 

In this section, we address the exact forms of the indices for the M5-brane giant gravitons by means of the giant graviton expansions 
of the Coulomb indices and the Higgs indices for the M2-brane SCFTs. 
Furthermore, we obtain from the expansions the exact form of the twisted indices of the M5-brane SCFTs 
and find that they also admit the giant graviton expansions with respect to the indices of the M2-giants as inverse transformations. 

The M5-brane (resp.~M2-brane) SCFTs on the boundary of $AdS_7$ (resp.~$AdS_4$) 
are essentially the same as the those on the wrapped giant M5-branes (resp.~giant M2-branes) in $AdS_4\times S^7$ (resp.~$AdS_7\times S^4$). 
The indices of these theories are related with each other under changes of fugacities. 

To identify the concrete transformation rules of the fugacities, let us first consider a spherical M5-brane giant graviton wrapping $S^5$ 
as an intersection of the plane $z_1=0$ with $S^7$ in the dual geometry $AdS_4\times S^7$ (see section \ref{sec_giants} for notation).
This configuration preserves half of the supersymmetry which commutes with $h-r_{12}$, and the remaining subalgebra contains as the bosonic subalgebra 
\begin{align}
\mathfrak{u}(1)_{h-r_{12}}\times \mathfrak{su}(2|4),\quad 
\mathfrak{su}(2|4)\supset \mathfrak{so}(3)_{j_{ij},i,j=1,2,3}\times \mathfrak{so}(6)_{r_{ab},a,b=3,\cdots,8}\times \mathfrak{u}(1)_{h-\frac{1}{2}r_{12}}
\label{3dsubalgebra}
\end{align}
generated by the operators in the subscripts. 
As argued in \cite{Arai:2020uwd}, there is an isomorphism between the symmetry algebra (\ref{3dsubalgebra}) preserved by the M5-brane giant graviton and the subalgebra of M5-brane SCFT preserved by a spherical M2-brane giant graviton configuration.
If we choose the M2-giant graviton as the one with $Z_{1}=0$ (see section \ref{sec_giants} for notation),\footnote{
Differenct choice of the M2-brane giant graviton at this step results in the same conclusion \eqref{M2M5dictionaryforz12} since the supersymmetric index of M5-brane SCFT \eqref{def_M5index} is symmetric under the exchange $u\rightarrow u^{-1}$.
}
the bosonic subalgebra of the remaining symmetry algebra in the M5-brane SCFT is
\begin{align}
\mathfrak{so}(2)_{H-R_{12}}\times \mathfrak{su}(4|2),\quad
\mathfrak{su}(4|2)\supset \mathfrak{su}(4)_{J_{ij},i,j=1,\cdots,6}\times \mathfrak{su}(2)_{R_{ab},a,b=3,4,5}\times \mathfrak{u}(1)_{H-2R_{12}}.
\label{6dsubalgebra}
\end{align}
By matching the subalgebras \eqref{3dsubalgebra} and \eqref{6dsubalgebra}, we find \cite{Arai:2020uwd}
\begin{align}
\label{match_subalg}
&h=\frac{H}{2}-\frac{3R_{12}}{2},\quad
j_{12}=R_{34},\quad
r_{12}=-R_{12},\quad
r_{34}=J_{12},\quad
r_{56}=J_{34},\quad
r_{78}=J_{56}.
\end{align}

The transformation of the fugacities for the M5-brane giant gravitons associated with $z_1=0$ can be obtained 
by substituting the generators (\ref{match_subalg}) to the supersymmetric index of the M2-brane SCFT \eqref{def_M2index} 
and 
rewriting it in terms of the supersymmetric indices of the M5-brane SCFT \eqref{def_M5index}. 
We find
\begin{align}
p=q^{\frac{1}{4}}t^{\frac{1}{3}}z^{-\frac{1}{3}},\quad
u=q^{-\frac{5}{8}}t^{\frac{1}{2}}z^{-\frac{1}{2}},\quad
y_1=t^{\frac{2}{3}}z^{\frac{1}{3}}x,\quad
y_2=t^{-\frac{4}{3}}z^{-\frac{2}{3}},\quad
y_3=t^{\frac{2}{3}}z^{\frac{1}{3}}x^{-1}.
\label{M2M5dictionaryforz12}
\end{align}
As reviewed in section \ref{sec_giants}, 
the M5-brane giant graviton preserving less supersymmetry can wrap more general supersymmetric cycles 
as intersections of the holomorphic surfaces (\ref{hol_giantM5}) in $\mathbb{C}^4$ with $S^7$ 
so that the holomorphic surfaces involve other zeros of the holomorphic equations with four variables.
The identification (\ref{M2M5dictionaryforz12}) of the parameters allows for a single sum giant graviton expansions of the M2-brane indices with less superseymmetry in such a way that the expansion coefficients are viewed as the indices of M5-brane giant gravitons with the same amount of supersymmetry. 

For the M5-brane giant graviton associated with $z_{3}=0$, 
we find the corresponding fugacity transformation
\begin{align}
p=q^{\frac{1}{4}}t^{\frac{1}{3}}z^{\frac{1}{3}},\quad
u=q^{-\frac{5}{8}}t^{\frac{1}{2}}z^{\frac{1}{2}},\quad
y_1=t^{\frac{2}{3}}z^{-\frac{1}{3}}x^{-1},\quad
y_2=t^{-\frac{4}{3}}z^{\frac{2}{3}},\quad
y_3=t^{\frac{2}{3}}z^{-\frac{1}{3}}x.
\label{M2M5dictionaryforz56}
\end{align}
This allows us the single sum giant graviton expansion of the M2-brane indices with respect to a different variable.

It is argued in \cite{Arai:2019xmp} that 
the leading terms in the giant graviton expansion of the supersymmetric indices for the SCFTs dual to the bulk $AdS$ geometries correspond to the ground state configuration 
characterized by the supersymmetric surface wrapped by the giant graviton. 
For the M5-brane giant graviton we have the following ground state contributions to the M2-brane index: 
\begin{align}
\begin{array}{cc}
S^5&\text{ground state contributions}\\ \hline
z_{1}=0&(q^{\frac{1}{4}}zt^{-1})^N\\ \hline
z_{2}=0&(q^{\frac{1}{4}}xt)^N\\ \hline
z_{3}=0&(q^{\frac{1}{4}}z^{-1}t^{-1})^N\\ \hline
z_{4}=0&(q^{\frac{1}{4}}x^{-1}t)^N\\
\end{array}. 
\end{align}
In particular, the coefficient of ($q^{\frac{1}{4}}zt^{-1})^{mN}$ (resp.~$(q^{\frac{1}{4}}z^{-1}t^{-1})^{mN}$) 
in the giant graviton expansion of the Coulomb index is expected to be given by the supersymmetric index of the fluctuation modes of $1/4$-BPS M5-brane giant gravitons of wrapping number $m$ associated with $z_1=0$ (resp.~$z_3=0$)
which will also be calculated as the 6d supesymmetric index \eqref{def_M5index} for $m$ M5-branes upon a proper identification of the Cartan generators.

Hence we conclude that the coefficients of $(q^{\frac{1}{4}}zt^{-1})^{mN}$ (resp.~$(q^{\frac{1}{4}}z^{-1}t^{-1})^{mN}$) in the giant graviton expansion 
of the supersymmetric indices of the M2-brane SCFT should be interpreted as the supersymmetric indices of $m$ M5-branes \eqref{def_M5index} with the fugacities $(u,y_1,y_2,y_3;p)$ substituted with \eqref{M2M5dictionaryforz12} (resp.~\eqref{M2M5dictionaryforz56}).
In particular, if we take the Coulomb limit \eqref{def_Cindex} with $x_1=\mathfrak{t}z^{-1}$ and $x_2=\mathfrak{t}z$, 
we find that the coefficients are respectively given by
\begin{align}
&\text{coefficient of }x_1^{mN}\rightarrow \sigma_1\Bigl[{\cal I}^{\text{M5 SCFT}}_\text{twist}[\mathbb{R}^5;m](x_1;x_2)\Bigr],\nonumber \\
&\text{coefficient of }x_2^{mN}\rightarrow \sigma_2\Bigl[{\cal I}^{\text{M5 SCFT}}_\text{twist}[\mathbb{R}^5;m](x_1;x_2)\Bigr],
\end{align}
with ${\cal I}^{\text{M5 SCFT}}_\text{twist}[\mathbb{R}^5;m](x_1;x_2)$ the twisted limit defined as \eqref{IM5twisted} and
\begin{align}
\label{var_change}
\sigma_1:\quad x_1\rightarrow x_1^{-1}, \qquad x_2\rightarrow x_1 x_2, \\
\label{var_change2}
\sigma_2:\quad x_1\rightarrow x_1x_2, \qquad x_2\rightarrow x_2^{-1}. 
\end{align}
Note that $\sigma_1\circ \sigma_1=\sigma_2\circ \sigma_2=\text{id}$.

For the orbifold M5-brane giant gravitons wrapping the $\mathbb{Z}_l$ homology cycle or the intersections of the holomorphic surfaces in $\mathbb{C}^4/\mathbb{Z}_l$ 
with the orbifold $S^7/\mathbb{Z}_l$, one finds mod $l$ wrapping number \cite{Aharony:2008ug,Lozano:2013ota}. 
Therefore the coefficients of $(q^{\frac{1}{4}}zt^{-1})^{lmN}$ (resp.~$(q^{\frac{1}{4}}z^{-1}t^{-1})^{lmN}$) in the giant graviton expansions  
of the Coulomb indices for the theories of M2-brane probing the orbifold will be identified with the supersymmetric indices of the $1/3$-BPS orbifold M5-brane giant gravitons. 

\subsection{Coulomb indices}
We examine the Coulomb indices of 3d $\mathcal{N}=4$ $U(N)$ ADHM theories with $l$ flavors \cite{deBoer:1996mp,deBoer:1996ck}.  
They enumerate the BPS local operators, which are responsible for the description of the geometries $\mathbb{C}^2/\mathbb{Z}_l$ probed by a stack of $N$ M2-branes. 
They also arise from the supersymmetric indices of the $U(N)_l\times U(N)_{-l}$ ABJM theories \cite{Aharony:2008ug} or the $U(N)_{l}\times U(N+l)_{-l}$ ABJ theories \cite{Gaiotto:2008sd,Hosomichi:2008jb,Aharony:2008gk} in either the Coulomb limit or the Higgs limit \cite{Hayashi:2022ldo}. 
They can be regarded as the refined Hilbert series for the $N$-th symmetric products of $\mathbb{C}^2/\mathbb{Z}_l$. 

\subsubsection{$\mathbb{C}^2$}
\label{sec_CoulombC2}
Let us consider the $U(N)$ ADHM theory with a single hypermultiplet in the fundamental representation that describes a stack of $N$ M2-branes in flat space $\mathbb{C}^4$. 
In this case, the theory has enhanced $\mathcal{N}=8$ supersymmetry and it is dual to the $U(N)_{1}\times U(1)_{-1}$ ABJM theory \cite{Kapustin:2010xq}. 
The gravity dual geometry is $AdS_4\times S^7$ in M-theory. 

Since the theory is self-mirror, the Coulomb index is equivalent to the Higgs index. 
The Coulomb index encodes the spectrum of the Coulomb branch operators describing the $\mathbb{C}^2$ probed by a stack of $N$ M2-branes. 
In terms of the fugacities $(x_1, x_2)$, it is given by \cite{Hayashi:2024jof}
\begin{align}
\label{ind_ADHMun1C}
\mathcal{I}^{\textrm{$U(N)$ ADHM-[1]}}_{C}[\mathbb{C}^4;N](x_1;x_2)
&=\sum_{
\begin{smallmatrix}
\lambda\\
l(\lambda)\le N
\end{smallmatrix}
}
\frac{(x_1 x_2)^{2|\lambda|} x_1^{2n(\lambda)-\sum_i {\lambda'_i}^2}}
{(x_1;x_1)_{N-l(\lambda)} \prod_{j\ge 1}(x_1;x_1)_{m_j(\lambda)}}, 
\end{align}
where the sum is taken over the partitions $\lambda$ 
of $|\lambda|$ whose lengths $l(\lambda)$ are not greater than $N$. 

While the formula (\ref{ind_ADHMun1C}) involves an infinite sum over the partitions, 
another useful formula as a finite sum over the partitions of $N$ can be derived by means of the Fermi-gas method \cite{Hayashi:2022ldo}. 
In the Fermi-gas method it is convenient to introduce the grand canonical index 
\begin{align}
\label{grand_C_C^2}
\Xi^{\mathbb{C}^2}_{C}(\kappa;x_1;x_2)
&=\sum_{N=0}^{\infty}\mathcal{I}^{\textrm{$U(N)$ ADHM-[1]}}_{C}[\mathbb{C}^2\times \mathbb{C}^2;N](x_1;x_2)\kappa^N. 
\end{align}
It can be evaluated as \cite{Hayashi:2022ldo}
\begin{align}
\label{grand_C_C^2_exact}
\Xi^{\mathbb{C}^2}_{C}(\kappa;x_1;x_2)&
=\exp \left[
\sum_{m=1}^{\infty}\frac{ \mathcal{I}^{\textrm{$U(1)$ ADHM-[1]}}_{C}[\mathbb{C}^2\times \mathbb{C}^2/\mathbb{Z}_2;1](x_1^m;x_2^m)}{m}\kappa^m \right]
\nonumber\\
&=\prod_{m=0}^{\infty}\frac{1}{1-\kappa x_1^m x_2^m}
\prod_{n=1}^{\infty}\frac{1}{(1-\kappa x_1^{m+n}x_2^{m}) (1-\kappa x_1^{m}x_2^{m+n})}. 
\end{align}
As the grand canonical index (\ref{grand_C_C^2_exact}) can be viewed as the Fredholm determinant, 
the index is given by
\begin{align}
\label{ind_ADHMun1C2}
\mathcal{I}^{\textrm{$U(N)$ ADHM-[1]}}_{C}[\mathbb{C}^4;N](x_1;x_2)
&=\sum_{\begin{smallmatrix}
\lambda\\
|\lambda|=N\\
\end{smallmatrix}}
\prod_{i=1}^{r}\frac{Z_{\lambda_i}[\mathbb{C}^4;N](x_1;x_2)^{m_i}}{\lambda_i^{m_i}m_i!}, 
\end{align}
where 
\begin{align}
Z_{i}[\mathbb{C}^4;N](x_1;x_2)&=\frac{1}{(1-x_1^i)(1-x_2^i)}
\end{align}
and $\lambda$ $=$ $(\lambda_1^{m_1}\lambda_2^{m_2}\cdots \lambda_r^{m_r})$ 
is a partition of integer $N$ with $\lambda_1>\lambda_2>\cdots >\lambda_{r}>\lambda_{r+1}=0$ and $\sum_{i=1}^{r}m_i \lambda_i=N$. 
When $N=1$ and $N=\infty$ the Coulomb index (\ref{ind_ADHMun1C}) or (\ref{ind_ADHMun1C2}) reproduces  
(\ref{Cindex_1M2}) and (\ref{Cindex_largeNM2}) respectively. 

The single sum giant graviton expansion of the Coulomb index (\ref{ind_ADHMun1C}) was proposed by Gaiotto and Lee \cite{Gaiotto:2021xce}, 
which takes the form 
\begin{align}
\label{GG1_ADHMun1C}
\frac{\mathcal{I}^{\textrm{$U(N)$ ADHM-[1]}}_{C}[\mathbb{C}^4;N](x_1;x_2)}
{\mathcal{I}^{\textrm{$U(\infty)$ ADHM-[1]}}_{C}[\mathbb{C}^4;N=\infty](x_1;x_2)}
&=\sum_{m=0}^{\infty}x_1^{Nm}\hat{F}_{m}^{\mathbb{C}^2}(x_1;x_2), 
\end{align}
where 
\begin{align}
\label{GGind_ADHMun1C}
\hat{F}_m^{\mathbb{C}^2}(x_1;x_2)&=\prod_{k=1}^{m}\prod_{n=0}^{\infty}
\frac{1}{1-x_1^{-k}x_2^{n}}
\end{align}
is identified with the index of the $1/4$-BPS M5-brane giant gravitons of wrapping number $m$. 
The single sum giant graviton expansion \eqref{GG1_ADHMun1C} can be checked by 
expanding the both sides in the fugacity $x_2$ and then comparing each of the coefficients of the terms in the expansion 
as a functions of the remaining fugacity $x_1$.  

For example, the index of a single $1/4$-BPS M5-brane giant graviton is
\begin{align}
\label{GGind_ADHMu11C}
\hat{F}_1^{\mathbb{C}^2}(x_1;x_2)&=
\prod_{n=0}^{\infty}\frac{1}{1-x_1^{-1}x_2^n}. 
\end{align}
Notice that it is simply obtained from the twisted index (\ref{Tindex_1M5}) for a single M5-brane under the transformation (\ref{var_change})
\begin{align}
\hat{F}_1^{\mathbb{C}^2}(x_1;x_2)&=
\sigma_1 \left[
\mathcal{I}^{\textrm{free tensor}}_{\textrm{twist}}[\mathbb{R}^5;N=1](x_1;x_2)
\right], 
\end{align}
as expected. 
When $N=1$, the giant graviton expansion (\ref{GG1_ADHMun1C}) can be stated as the following identity
\begin{align}
&
\prod_{n=1}^{\infty}(1-x_1^nx_2^n)
\prod_{n=2}^{\infty}(1-x_1^n)(1-x_2^n)
\prod_{m=1}^{\infty}\prod_{n=1}^{\infty}
(1-x_1^{m+n}x_2^{m})(1-x_1^{m}x_2^{m+n})
\nonumber\\
&=\sum_{m=0}^{\infty}
\frac{x_1^m}{(x_1^{-1};x_2)_{\infty}(x_1^{-2};x_2)_{\infty}\cdots (x_1^{-m};x_2)_{\infty}}. 
\end{align}

Upon the variable change (\ref{var_change}), 
we obtain the closed-form expression for the twisted index for 6d $(2,0)$ theory describing $N$ coincident M5-branes 
\begin{align}
\label{6dind_1C}
\mathcal{I}^{\textrm{6d $(2,0)$}}_{\textrm{twist}}[\mathbb{R}^5;N](x_1;x_2)
&=\sigma_1 \Bigl[ \hat{F}_{N}^{\mathbb{C}^2}(x_1;x_2) \Bigr]
\nonumber\\
&=\prod_{k=1}^{N}\prod_{n=0}^{\infty}\frac{1}{1-x_1^{n+k}x_2^n}. 
\end{align}

When we set $x_1$ to $p$ and $x_2$ to $1$, 
the dual index (\ref{6dind_1C}) becomes the unrefined index of the M5-brane SCFT
\begin{align}
\mathcal{I}^{\textrm{6d $(2,0)$}}_{\textrm{unref}}[\mathbb{R}^5;N](p)
&=\prod_{k=1}^{N}\prod_{n=0}^{\infty}\frac{1}{1-p^{n+k}}, 
\end{align}
which agrees with the results \cite{Kim:2012ava,Beem:2014kka}, as anticipated. 
It is identified with the vacuum character $\chi_{\mathcal{W}(\mathfrak{gl}(N))}(p)$ of the W-algebra $\mathcal{W}(\mathfrak{gl}(N))$ 
as the protected chiral algebra of 6d $(2,0)$ theory of type $\mathfrak{g}$ is conjectured to be isomorphic to the W-algebra $\mathcal{W}(\mathfrak{g})$ 
\cite{Beem:2014kka}. 

Here we observe that 
the M5-brane index (\ref{6dind_1C}) also has the giant graviton expansion
\begin{align}
\label{GG6d_1C}
\frac{\mathcal{I}^{\textrm{6d $(2,0)$}}_{\textrm{twist}}[\mathbb{R}^5;N](x_1;x_2)}
{\mathcal{I}^{\textrm{6d $(2,0)$}}_{\textrm{twist}}[\mathbb{R}^5;N=\infty](x_1;x_2)}
&=\sum_{m=0}^{\infty}
x_1^{Nm}\hat{G}_{m}^{\mathbb{C}^2}(x_1;x_2), 
\end{align}
where $\hat{G}_{m}^{\mathbb{C}^2}$ can be viewed as the index for M2-giants of wrapping number $m$. 
It is simply obtained from the Coulomb index (\ref{ind_ADHMun1C}) for the $U(m)$ ADHM theory with one flavor under the transformation (\ref{var_change})
\begin{align}
\hat{G}_{m}^{\mathbb{C}^2}(x_1;x_2)&=\sigma_1 \Bigl[ \mathcal{I}^{\textrm{$U(m)$ ADHM-[1]}}_{C}[\mathbb{C}^4;N](x_1;x_2) \Bigr]. 
\end{align}

It is instructive to take the specialization of the indices. 
We note that the expansion (\ref{GG6d_1C}) still holds in the unrefined limit (\ref{IM5unref}), 
for which the M5-brane indices become the vacuum characters of the W-algebra and 
the expansion coefficients are given by the Coulomb index (\ref{ind_ADHMun1C}) 
with the specialization $x_1$ $\rightarrow$ $p^{-1}$, $x_2$ $\rightarrow$ $p$
\begin{align}
\label{GG6d_1C_unref}
\frac{\mathcal{I}^{\textrm{6d $(2,0)$}}_{\textrm{unref}}[\mathbb{R}^5;N](p)}
{\mathcal{I}^{\textrm{6d $(2,0)$}}_{\textrm{unref}}[\mathbb{R}^5;N=\infty](p)}
&=\sum_{m=0}^{\infty}
p^{Nm}\hat{G}_{m}^{\mathbb{C}^2}(p), 
\end{align}
where $\hat{G}_{m}^{\mathbb{C}^2}(p)$ $=$ $\mathcal{I}^{\textrm{$U(N)$ ADHM-[1]}}_{C}[\mathbb{C}^4;N](p^{-1};p)$. 
While the leading coefficients in the giant graviton expansion of the unrefined index of the 6d $(2,0)$ theory 
describing a stack of $N$ M5-branes was addressed in \cite{Beccaria:2023sph}, 
our formula (\ref{GG6d_1C_unref}) now gives rise to all the expansion coefficients 
in terms of the Coulomb/Higgs indices of the ADHM theory with one flavor. 
For example, 
\begin{align}
\hat{G}_{1}^{\mathbb{C}^2}(p)&=-\frac{p}{(1-p)^2}, \\
\hat{G}_{2}^{\mathbb{C}^2}(p)&=\frac{2p^3}{(1-q)^2(1-q^2)^2}, \\
\hat{G}_{3}^{\mathbb{C}^2}(p)&=-\frac{p^5(1+4p+p^2)}{(1-p)^2(1-p^2)^2(1-p^3)^2}, \\
\hat{G}_{4}^{\mathbb{C}^2}(p)&=\frac{p^8(3+4p+10p^2+4p^3+3p^4)}{(1-p)^2(1-p^2)^2(1-p^3)^2(1-p^4)^2}, \\
\hat{G}_{5}^{\mathbb{C}^2}(p)&=-\frac{p^{11}(3+8p+15p^2+20p^3+28p^4+20p^5+15p^6+8p^7+3p^8)}{(1-p)^2(1-p^2)^2(1-p^3)^2(1-p^4)^2(1-p^5)^2}. 
\end{align}

Next we turn to the half-BPS indices. 
When we take the half-BPS limit of (\ref{ind_ADHMun1C}) (or equivalently (\ref{ind_ADHMun1C2})) and (\ref{6dind_1C}), 
we get
\begin{align}
\label{1/2M2}
\mathcal{I}^{\textrm{$U(N)$ ADHM-[1]}}_{\textrm{$\frac12$BPS}}[\mathbb{C}^4;N](x_1)
&=\prod_{n=1}^{N}\frac{1}{1-x_1^n}, \\
\label{1/2M5}
\mathcal{I}^{\textrm{6d $(2,0)$}}_{\textrm{$\frac12$BPS}}[\mathbb{R}^5;N](x_1)
&=\prod_{n=1}^{N}\frac{1}{1-x_1^n}. 
\end{align}
The half-BPS index for the theory of $N$ M2-branes probing $\mathbb{C}^4$ 
and that for the theory of $N$ M5-branes probing $\mathbb{R}^5$ have the same form. 
The giant graviton expansion of the half-BPS indices are given by
\begin{align}
\label{1/2M2gg}
\frac{\mathcal{I}^{\textrm{$U(N)$ ADHM-[1]}}_{\textrm{$\frac12$BPS}}[\mathbb{C}^4;N](x_1)}
{\mathcal{I}^{\textrm{$U(\infty)$ ADHM-[1]}}_{\textrm{$\frac12$BPS}}[\mathbb{C}^4;N=\infty](x_1)}
&=\sum_{m=0}^{\infty}x_1^{Nm}\hat{f}_{m}^{\mathbb{C}^2}(x_1),  \\
\label{1/2M5gg}
\frac{\mathcal{I}^{\textrm{6d $(2,0)$}}_{\textrm{$\frac12$BPS}}[\mathbb{R}^5;N](x_1)}
{\mathcal{I}^{\textrm{6d $(2,0)$}}_{\textrm{$\frac12$BPS}}[\mathbb{R}^5;N=\infty](x_1)}
&=\sum_{m=0}^{\infty}x_1^{Nm}\hat{g}_{m}^{\mathbb{C}^2}(x_1),
\end{align}
where 
\begin{align}
\label{1/2M5ginat_C^2}
\hat{f}_m^{\mathbb{C}^2}(x_1)&=\frac{(-1)^m x_1^{\frac{m(m+1)}{2}}}{(x_1;x_1)_{m}}
=\mathcal{I}^{\textrm{6d $(2,0)$}}_{\textrm{$\frac12$BPS}}[\mathbb{R}^5;m](x_1^{-1})
\end{align}
and 
\begin{align}
\label{1/2M2ginat_C^2}
\hat{g}_m^{\mathbb{C}^2}(x_1)&=\frac{(-1)^m x_1^{\frac{m(m+1)}{2}}}{(x_1;x_1)_{m}}
=\mathcal{I}^{\textrm{$U(m)$ ADHM-[1]}}_{\textrm{$\frac12$BPS}}[\mathbb{C}^4;m](x_1^{-1})
\end{align}
are the index for the half-BPS M5-brane giant gravitons of wrapping number $m$ and that for the half-BPS M2-brane giant gravitons of wrapping number $m$ respectively. 
The expressions (\ref{1/2M5ginat_C^2}) and (\ref{1/2M2ginat_C^2}) are exactly same as the index 
for the half-BPS D3-brane giant gravitons \cite{Gaiotto:2021xce,Lee:2023iil,Eleftheriou:2023jxr,Deddo:2024liu}. 
This is consistent with the semiclassical analysis \cite{Eleftheriou:2023jxr} of the half-BPS giant gravitons based on the brane actions 
where the spectra are addressed from the Lagrangian for quadratic fluctuations of a spherical D3-brane, M2-brane and M5-brane in the $AdS_{m}\times S^{n}$. 

As the indices contain two variables corresponding to the two planes $z_1=0$ and $z_3=0$, let us now turn to the double sum expansions. For each pair $(m_1,m_2)$ of wrapping numbers the expansion coefficients will consist of the indices (\ref{GGind_ADHMun1C}) for the M5-brane giant gravitons. Besides, the configuration will generally involve M2-branes stretched between them whose contributions will be independent of the background flux $N$. 
Assuming that the both two variables $x_1$ and $x_2$ are small enough within a unit torus, we find that these indices admit the double sum expansions in a similar manner as for the expansion of the Schur index \cite{Arai:2020qaj}. We find that
\begin{align}
\frac{\mathcal{I}^{\textrm{$U(N)$ ADHM-[1]}}_{C}[\mathbb{C}^4;N](x_1;x_2)}
{\mathcal{I}^{\textrm{$U(\infty)$ ADHM-[1]}}_{C}[\mathbb{C}^4;N=\infty](x_1;x_2)}
&=\sum_{m_1=0}^{\infty}\sum_{m_2=0}^{\infty}
x_1^{Nm_1} x_2^{Nm_2 } (x_1 x_2)^{m_1 m_2}
\nonumber\\
&\times 
\hat{F}_{m_1}^{\mathbb{C}^2}(x_1;x_2)
\hat{F}_{m_2}^{\mathbb{C}^2}(x_2;x_1).
\end{align}
The factor $(x_1x_2)^{m_1 m_2}$ will be interpreted as the contributions from the M2-branes suspended between the M5-brane giant gravitons.

Analogously, we find that the twisted index can be expanded a double sum of the form
\begin{align}
\frac{\mathcal{I}^{\textrm{6d $(2,0)$}}_{\textrm{twist}}[\mathbb{R}^5;N](x_1;x_2)}
{\mathcal{I}^{\textrm{6d $(2,0)$}}_{\textrm{twist}}[\mathbb{R}^5;N=\infty](x_1;x_2)}
&=\sum_{m_1=0}^{\infty}\sum_{m_2=0}^{\infty}
x_1^{Nm_1}x_2^{Nm_2}(x_1 x_2)^{m_1 m_2}
\nonumber\\
&\times 
\hat{G}_{m_1}^{\mathbb{C}^2}(x_1;x_2)
\hat{G}_{m_2}^{\mathbb{C}^2}(x_2;x_1). 
\end{align}
We note that the expansions are not unique and there will be other form of the expansions depending on the regions of the expanded fugacities. It would be interesting to figure out the full structure of the admissible expansions and their  physical meaning, such as wall-crossing phenomena and the associated brane configurations.

\subsubsection{$\mathbb{C}^2/\mathbb{Z}_2$}
Next consider the case with $l=2$. 
The $U(N)$ ADHM theory preserves $\mathcal{N}=4$ supersymmetry and the gravity dual geometry is $AdS_4\times S^7/\mathbb{Z}_2$ in M-theory. 
The Coulomb index is the refined Hilbert series for the $N$-th symmetric product of $\mathbb{C}^2/\mathbb{Z}_2$. 
It also agrees with the Coulomb or Higgs index of the $U(N)_{2}\times U(N)_{-2}$ ABJM theory \cite{Hayashi:2022ldo}.  

Let us define the grand canonical Coulomb index of the $U(N)$ ADHM theory with two flavors by
\begin{align}
\Xi^{\mathbb{C}^2/\mathbb{Z}_2}_{C}(\kappa;x_1;x_2)
&=\sum_{N=0}^{\infty}\mathcal{I}^{\textrm{$U(N)$ ADHM-[2]}}_{C}[\mathbb{C}^2\times \mathbb{C}^2/\mathbb{Z}_2;N](x_1;x_2) \kappa^N. 
\end{align}
It can be computed as \cite{Hayashi:2022ldo}
\begin{align}
\Xi^{\mathbb{C}^2/\mathbb{Z}_2}_{C}(\kappa;x_1;x_2)&
=\exp \left[
\sum_{m=1}^{\infty}\frac{ \mathcal{I}^{\textrm{$U(1)$ ADHM-[2]}}_{C}[\mathbb{C}^2\times \mathbb{C}^2/\mathbb{Z}_2;1](x_1^m;x_2^m)}{m}\kappa^m \right]
\nonumber\\
&=\prod_{m=0}^{\infty}\frac{1}{1-\kappa x_1^m x_2^m}
\prod_{n=1}^{\infty}\frac{1}{(1-\kappa x_1^{m+2n}x_2^{m}) (1-\kappa x_1^{m}x_2^{m+2n})}, 
\end{align}
where 
\begin{align}
\mathcal{I}^{\textrm{$U(1)$ ADHM-[2]}}_{C}[\mathbb{C}^2\times \mathbb{C}^2/\mathbb{Z}_2;1](x_1;x_2)
&=\frac{1-x_1^2x_2^2}{(1-x_1x_2)(1-x_1^2)(1-x_2^2)}
\end{align}
is the Coulomb index for the Abelian theory with $N=1$. 
The canonical index reads
\begin{align}
\label{ind_ADHMun2C_C^2/Z_2}
&
\mathcal{I}^{\textrm{$U(N)$ ADHM-[2]}}_{C}[\mathbb{C}^2\times \mathbb{C}^2/\mathbb{Z}_{2};N](x_1;x_2)
\nonumber\\
&=\sum_{\begin{smallmatrix}
\lambda\\
|\lambda|=N\\
\end{smallmatrix}}
\prod_{i=1}^{r}\frac{Z_{\lambda_i}[\mathbb{C}^2\times \mathbb{C}^2/\mathbb{Z}_2;N](x_1;x_2)^{m_i}}{\lambda_i^{m_i}m_i!}, 
\end{align}
where 
\begin{align}
Z_{i}[\mathbb{C}^2\times \mathbb{C}^2/\mathbb{Z}_2;N](x_1;x_2)
&=\frac{1+x_1^i x_2^i}{(1-x_1^{2i})(1-x_2^{2i})}. 
\end{align}
In the large $N$ limit the Coulomb index is given by \cite{Hayashi:2022ldo}
\begin{align}
&\mathcal{I}^{\textrm{$U(\infty)$ ADHM-[2]}}_{C}[\mathbb{C}^2\times \mathbb{C}^2/\mathbb{Z}_2;\infty](x_1;x_2)
\nonumber\\
&=\prod_{n=1}^{\infty}\frac{1}{1-x_1^nx_2^n}
\prod_{m=0}^{\infty}\prod_{n=1}^{\infty}\frac{1}{(1-x_1^{m+2n}x_2^{m})(1-x_1^{m}x_2^{m+2n})}. 
\end{align}

We find that the Coulomb index of the $U(N)$ ADHM theory with two flavors admits the following single sum giant graviton expansion: 
\begin{align}
\label{GGC_C^2/Z_2}
\frac{\mathcal{I}^{\textrm{$U(N)$ ADHM-[2]}}_{C}[\mathbb{C}^2\times \mathbb{C}^2/\mathbb{Z}_2;N](x_1;x_2)}
{\mathcal{I}^{\textrm{$U(\infty)$ ADHM-[2]}}_{C}[\mathbb{C}^2\times \mathbb{C}^2/\mathbb{Z}_2;N=\infty](x_1;x_2)}
&=\sum_{m=0}^{\infty}x_1^{2Nm}\hat{F}_{m}^{\mathbb{C}^2/\mathbb{Z}_2}(x_1;x_2), 
\end{align}
where 
\begin{align}
\label{GGmind_C^2/Z_2}
\hat{F}_{m}^{\mathbb{C}^2/\mathbb{Z}_2}(x_1;x_2)
&=\prod_{k=1}^{m}\prod_{n=0}^{\infty}\frac{1}{(1-x_1^{-2k}x_2^{2n})(1-x_1^{-2k+1}x_2^{2n+1})} 
\end{align}
will be identified with the $1/3$-BPS index for the M5-brane giant gravitons wrapping a $5$-cycle in the $S^7/\mathbb{Z}_2$. 
Unlike the case with one flavor, the expansion (\ref{GGC_C^2/Z_2}) has contributions only from the giants of even wrapping numbers $2m$ with $m=0,1,\cdots$. 

For $m=1$ corresponding to the M5-brane giant graviton with wrapping number $2$, 
the single particle index is
\begin{align}
\label{GG1ind_C^2/Z_2}
\hat{i}_{m=1}^{\mathbb{C}^2/\mathbb{Z}_2}(x_1;x_2)
&=\frac{x_1^{-1}x_2}{1-x_2^2}+\frac{x_1^{-2}}{1-x_2^2}. 
\end{align}

The twisted index for $N$ M5-branes wrapping $S^5/\mathbb{Z}_2$ will be obtained under the variable change (\ref{var_change}). 
We get
\begin{align}
\label{6dind_2C}
\mathcal{I}^{\textrm{6d $(2,0)$ $\mathbb{Z}_2$}}_{\textrm{twist}}[\mathbb{R}^5;N](x_1;x_2)
&=\sigma_1 \Bigl[ \hat{F}_{N}^{\mathbb{C}^2/\mathbb{Z}_2}(x_1;x_2) \Bigr]
\nonumber\\
&=\prod_{k=1}^{N}\prod_{n=0}^{\infty}\frac{1}{(1-x_1^{2n+2k}x_2^{2n})(1-x_1^{2n+2k}x_2^{2n+1})}. 
\end{align}
For example for $N=1$, we have
\begin{align}
\mathcal{I}^{\textrm{6d $(2,0)$ $\mathbb{Z}_2$}}_{\textrm{twist}}[\mathbb{R}^5;N=1](x_1;x_2)
&=\prod_{n=0}^{\infty}
\frac{1}{(1-x_1^{2n+2} x_2^{2n}) (1-x_1^{2n+2}x_2^{2n+1})}
\end{align}
and the corresponding single particle twisted index reads
\begin{align}
\label{6dsind_2C}
i^{\textrm{6d $(2,0)$ $\mathbb{Z}_2$}}_{\textrm{twist}}[\mathbb{R}^5;N=1](x_1;x_2)
&=\frac{x_1^2 x_2}{1-x_1^2 x_2^2}(1+x_2^{-1}). 
\end{align}
We observe that the single particle index (\ref{6dsind_2C}) is given by a sum of two parts 
with the common factor $x_1^2 x_2/(1-x_1^2 x_2^2)$ corresponding to the $\mathbb{Z}_2$-projection of the twisted index for the 6d $(2,0)$ Abelian theory 
(see (\ref{sgTindex_1M5}) for $l=2$). 
They are distinguished by a phase factor $x_2^{-1}$. 

Consider the giant graviton expansion for the twisted index (\ref{6dind_2C}) of the form 
\begin{align}
\label{GG6d_2C}
\frac{\mathcal{I}^{\textrm{6d $(2,0)$ $\mathbb{Z}_2$}}_{\textrm{twist}}[\mathbb{R}^5;N](x_1;x_2)}
{\mathcal{I}^{\textrm{6d $(2,0)$ $\mathbb{Z}_2$}}_{\textrm{twist}}[\mathbb{R}^5;N=\infty](x_1;x_2)}
&=\sum_{m=0}^{\infty}
x_1^{2Nm} \hat{G}_{m}^{\mathbb{C}^2/\mathbb{Z}_2}(x_1;x_2). 
\end{align}
We have confirmed that 
the index of the M2-giant gravitons is beautifully obtained from the ADHM index with two flavors under the variable change (\ref{var_change})
\begin{align}
\hat{G}_{m}^{\mathbb{C}^2/\mathbb{Z}_2}(x_1;x_2)
&=\sigma_1 \Bigl[ \mathcal{I}^{\textrm{$U(m)$ ADHM-[2]}}_{C}[\mathbb{C}^2\times \mathbb{C}^2/\mathbb{Z}_2;\infty](x_1;x_2) \Bigr].
\end{align}

Note that the expansion (\ref{GG6d_2C}) has the unrefined limit 
in such a way that the expansion coefficients are given by the specialized Coulomb index (\ref{ind_ADHMun2C_C^2/Z_2}) 
with $x_2$ $\rightarrow$ $x_1^{-1}$. 
In the half-BPS limit, we have 
\begin{align}
\label{1/2M2Z2}
\mathcal{I}^{\textrm{$U(N)$ ADHM-[2]}}_{\textrm{$\frac12$BPS},C}[\mathbb{C}^2\times \mathbb{C}^2/\mathbb{Z}_{2};N](x_1)
&=\prod_{n=1}^{N}\frac{1}{1-x_1^{2n}}, \\
\label{1/2M5Z2}
\mathcal{I}^{\textrm{6d $(2,0)$ $\mathbb{Z}_2$}}_{\textrm{$\frac12$BPS}}[\mathbb{R}^5;N](x_1)
&=\prod_{n=1}^{N}\frac{1}{1-x_1^{2n}}. 
\end{align}
Again the half-BPS index for the M2-brane SCFT is equal to that for the M5-brane SCFT. 
It is straightforward to show that they have the giant graviton expansions of the form
\begin{align}
\label{1/2M2Z2gg}
\frac{\mathcal{I}^{\textrm{$U(N)$ ADHM-[2]}}_{\textrm{$\frac12$BPS},C}[\mathbb{C}^2\times \mathbb{C}^2/\mathbb{Z}_2;N](x_1)}
{\mathcal{I}^{\textrm{$U(\infty)$ ADHM-[2]}}_{\textrm{$\frac12$BPS},C}[\mathbb{C}^2\times \mathbb{C}^2/\mathbb{Z}_2;N=\infty](x_1)}
&=\sum_{m=0}^{\infty}x_1^{2Nm}\hat{f}_{m}^{\mathbb{C}^2/\mathbb{Z}_2}(x_1), \\
\label{1/2M5Z2gg}
\frac{\mathcal{I}^{\textrm{6d $(2,0)$ $\mathbb{Z}_2$}}_{\textrm{$\frac12$BPS}}[\mathbb{R}^5;N](x_1)}
{\mathcal{I}^{\textrm{6d $(2,0)$ $\mathbb{Z}_2$}}_{\textrm{$\frac12$BPS}}[\mathbb{R}^5;N=\infty](x_1)}
&=\sum_{m=0}^{\infty}x_1^{2Nm}\hat{g}_{m}^{\mathbb{C}^2/\mathbb{Z}_2}(x_1), 
\end{align}
where 
\begin{align}
\hat{f}_{m}^{\mathbb{C}^2/\mathbb{Z}_2}(x_1)&=\frac{(-1)^m x_1^{m(m+1)}}{(x_1^2;x_1^2)_{m}}
=\mathcal{I}^{\textrm{6d $(2,0)$ $\mathbb{Z}_2$}}_{\textrm{$\frac12$BPS}}[\mathbb{R}^5;m](x_1^{-1})
\end{align}
is identified with the index for the half-BPS M5-brane giants of even wrapping number $2m$ in $S^7/\mathbb{Z}_2$ 
and
\begin{align}
\hat{g}_{m}^{\mathbb{C}^2/\mathbb{Z}_2}(x_1)&=\frac{(-1)^m x_1^{m(m+1)}}{(x_1^2;x_1^2)_{m}}
=\mathcal{I}^{\textrm{$U(m)$ ADHM-[2]}}_{\textrm{$\frac12$BPS},C}[\mathbb{C}^2\times \mathbb{C}^2/\mathbb{Z}_2;m](x_1^{-1})
\end{align}
can be viewed as the index for the half-BPS M2-brane giants of even wrapping number $2m$ in $S^4/\mathbb{Z}_2$ as the half-BPS index (\ref{1/2M5Z2}) is equal to the half-BPS index for the M5-branes probing the orbifold geometry $\mathbb{R}^5/\mathbb{Z}_2$.

In addition, we find that the indices have the double sum giant graviton expansions
\begin{align}
\frac{\mathcal{I}^{\textrm{$U(N)$ ADHM-[2]}}_{C}[\mathbb{C}^2\times \mathbb{C}^2/\mathbb{Z}_2;\infty](x_1;x_2)}
{\mathcal{I}^{\textrm{$U(\infty)$ ADHM-[2]}}_{C}[\mathbb{C}^2\times \mathbb{C}^2/\mathbb{Z}_2;\infty](x_1;x_2)}
&=\sum_{m_1=0}^{\infty}
\sum_{m_2=0}^{\infty}
x_1^{2Nm_1}
x_2^{2Nm_2}
(x_1 x_2)^{m_1 m_2}
\nonumber\\
&\times 
\hat{F}_{m_1}^{\mathbb{C}^2/\mathbb{Z}_2}(x_1;x_2)
\hat{F}_{m_2}^{\mathbb{C}^2/\mathbb{Z}_2}(x_2;x_1) 
\end{align}
and 
\begin{align}
\frac{\mathcal{I}^{\textrm{6d $(2,0)$ $\mathbb{Z}_2$}}_{\textrm{twist}}[\mathbb{R}^5;N](x_1;x_2)}
{\mathcal{I}^{\textrm{6d $(2,0)$ $\mathbb{Z}_2$}}_{\textrm{twist}}[\mathbb{R}^5;N=\infty](x_1;x_2)}
&=\sum_{m_1=0}^{\infty}
\sum_{m_2=0}^{\infty}
x_1^{2Nm_1}
x_2^{2Nm_2}
(x_1 x_2)^{m_1 m_2}
\nonumber\\
&\times 
\hat{G}_{m_1}^{\mathbb{C}^2/\mathbb{Z}_2}(x_1;x_2)
\hat{G}_{m_2}^{\mathbb{C}^2/\mathbb{Z}_2}(x_2;x_1). 
\end{align}

\subsubsection{$\mathbb{C}^2/\mathbb{Z}_l$}
Now consider general $U(N)$ ADHM theory with $l$ fundamental hypermultiplets. 
The theory describes a stack of $N$ coincident M2-branes probing $\mathbb{C}^2\times \mathbb{C}^2/\mathbb{Z}_l$. 
The gravity dual geometry is $AdS_4\times S^7/\mathbb{Z}_{l}$ in M-theory. 
The Coulomb index is the refined Hilbert series for the $N$-th symmetric product of $\mathbb{C}^2/\mathbb{Z}_l$. 
Since the theory is mirror to the $U(N)^{\otimes l}$ necklace quiver gauge theory \cite{deBoer:1996ck}, it is equal to the Higgs index of its mirror theory \cite{Cremonesi:2013lqa,Hayashi:2022ldo}. 
In addition, since the Coulomb or Higgs branch operators in the $U(N)_{l}\times U(N)_{-l}$ ABJM theory 
are responsible for the description of the geometry $\mathbb{C}^2/\mathbb{Z}_l$ $\subset$ $\mathbb{C}^4/\mathbb{Z}_l$ probed by $N$ M2-branes,  
it agrees with the Coulomb or Higgs index for the ABJM theory \cite{Hayashi:2022ldo}.

The grand canonical Coulomb index is given by \cite{Hayashi:2022ldo}
\begin{align}
\Xi^{\mathbb{C}^2/\mathbb{Z}_l}_{C}(\kappa;x_1;x_2)
&=\sum_{N=0}^{\infty}\mathcal{I}^{\textrm{$U(N)$ ADHM-[$l$]}}_{C}[\mathbb{C}^2\times \mathbb{C}^2/\mathbb{Z}_l;N](x_1;x_2) \kappa^N
\nonumber\\
&=\exp \left[
\sum_{m=1}^{\infty}\frac{ \mathcal{I}^{\textrm{$U(1)$ ADHM-[$l$]}}_{C}[\mathbb{C}^2\times \mathbb{C}^2/\mathbb{Z}_l;1](x_1^m;x_2^m)}{m}\kappa^m \right]
\nonumber\\
&=\prod_{m=0}^{\infty}\frac{1}{1-\kappa x_1^m x_2^m}
\prod_{n=1}^{\infty}\frac{1}{(1-\kappa x_1^{m+ln}x_2^{m}) (1-\kappa x_1^{m}x_2^{m+ln})}, 
\label{gcUNlADHMC}
\end{align}
where 
\begin{align}
\mathcal{I}^{\textrm{$U(1)$ ADHM-[$l$]}}_{C}[\mathbb{C}^2\times \mathbb{C}^2/\mathbb{Z}_l;1](x_1;x_2)
&=\frac{1-x_1^l x_2^l}{(1-x_1x_2)(1-x_1^l)(1-x_2^l)}
\end{align}
is the Coulomb index for the Abelian ADHM theory with $l$ flavors. 
The canonical index is given by
\begin{align}
\label{ind_ADHMunlC_C^2/Z_l}
&
\mathcal{I}^{\textrm{$U(N)$ ADHM-[$l$]}}_{C}[\mathbb{C}^2\times \mathbb{C}^2/\mathbb{Z}_{l};N](x_1;x_2)
\nonumber\\
&=\sum_{\begin{smallmatrix}
\lambda\\
|\lambda|=N\\
\end{smallmatrix}}
\prod_{i=1}^{r}\frac{Z_{\lambda_i}[\mathbb{C}^2\times \mathbb{C}^2/\mathbb{Z}_l;N](x_1;x_2)^{m_i}}{\lambda_i^{m_i}m_i!}, 
\end{align}
where 
\begin{align}
Z_{i}[\mathbb{C}^2\times \mathbb{C}^2/\mathbb{Z}_l;N](x_1;x_2)
&=\frac{(1-x_1^{li}x_2^{li})}{(1-x_1^{i} x_2^{i})(1-x_1^{li})(1-x_2^{li})}. 
\end{align}
In the large $N$ limit one finds \cite{Hayashi:2022ldo}
\begin{align}
&\mathcal{I}^{\textrm{$U(\infty)$ ADHM-[$l$]}}_{C}[\mathbb{C}^2\times \mathbb{C}^2/\mathbb{Z}_l;N=\infty](x_1;x_2)
\nonumber\\
&=\prod_{n=1}^{\infty}\frac{1}{1-x_1^nx_2^n}
\prod_{m=0}^{\infty}\prod_{n=1}^{\infty}\frac{1}{(1-x_1^{m+ln}x_2^{m})(1-x_1^{m}x_2^{m+ln})}. 
\end{align}

The finite $N$ correction of the ADHM Coulomb index will be controlled by the M5-giant gravitons with mod $l$ wrapping number. 
We propose the giant graviton expansion of the Coulomb index of the $U(N)$ ADHM theory with $l$ flavors of the form
\begin{align}
\frac{\mathcal{I}^{\textrm{$U(N)$ ADHM-[$l$]}}_{C}[\mathbb{C}^2\times \mathbb{C}^2/\mathbb{Z}_l;N](x_1;x_2)}
{\mathcal{I}^{\textrm{$U(\infty)$ ADHM-[$l$]}}_{C}[\mathbb{C}^2\times \mathbb{C}^2/\mathbb{Z}_l;N=\infty](x_1;x_2)}
&=\sum_{m=0}^{\infty}x_1^{lNm}\hat{F}_{m}^{\mathbb{C}^2/\mathbb{Z}_l}(x_1;x_2). 
\label{UNlCsinglesumgg}
\end{align}

The expansion coefficient in (\ref{UNlCsinglesumgg}) is identified with the index of the $1/3$-BPS orbifold M5-brane giant gravitons of wrapping number $ml$ 
supported on a supersymmetric $5$-cycle as an intersection of holomorphic surface defined by $f(z_1,z_3)=0$ in $\mathbb{C}^4/\mathbb{Z}_l$ with the $S^7/\mathbb{Z}_l$. 
We find that it is given by
\begin{align}
\hat{F}_m^{\mathbb{C}^2/\mathbb{Z}_l}(x_1;x_2)
&=\prod_{k=1}^{m}\prod_{i=0}^{l-1}\prod_{n=0}^{\infty}\frac{1}{(1-x_1^{-lk+i}x_2^{ln+i})}. 
\end{align}

Furthermore, the twisted index for 6d $(2,0)$ theory on $S^1\times S^5/\mathbb{Z}_l$ will be obtained upon the transformation (\ref{var_change}). 
We find 
\begin{align}
\label{6dind_lC}
\mathcal{I}^{\textrm{6d $(2,0)$ $\mathbb{Z}_l$}}_{\textrm{twist}}[\mathbb{R}^5;N](x_1;x_2)
&=\sigma_1 \Bigl[ \hat{F}_{N}^{\mathbb{C}^2/\mathbb{Z}_l}(x_1;x_2) \Bigr]
\nonumber\\
&=\prod_{k=1}^{N}\prod_{i=0}^{l-1}\prod_{n=0}^{\infty}\frac{1}{(1-x_1^{ln+lk}x_2^{ln+i})}. 
\end{align}
For example, for $N=1$, one finds from (\ref{6dind_lC}) the single particle twisted index for the $(2,0)$ theory of the free tensor multiplet on $S^1\times S^5/\mathbb{Z}_l$
\begin{align}
\label{6dsind_lC}
i^{\textrm{6d $(2,0)$ $\mathbb{Z}_l$}}_{\textrm{twist}}[\mathbb{R}^5;N=1](x_1;x_2)
&=
\frac{x_1^{l}x_2^{l-1}}{1-x_1^l x_2^l}
\left(
1+x_2^{-1}+x_2^{-2}+\cdots+x_2^{-l+1}
\right). 
\end{align}
We see that there are $l$ terms in (\ref{6dsind_lC}) corresponding to $l$ degenerate vacua. 
Each has the common factor $x_1^{l}x_2^{l-1}/(1-x_1^lx_2^l)$ 
that is equivalent to (\ref{sgTindex_1M5}), the $\mathbb{Z}_l$-projection of the twisted index. 
Each of them is distinguished by the phase factor $x_2^{-1}$, 
which is the fugacity corresponding to the squashing parameter of $S^5$. 

Conversely, we propose the giant graviton expansion of the twisted index (\ref{6dind_lC}) in the following form: 
\begin{align}
\label{GG6d_lC}
\frac{\mathcal{I}^{\textrm{6d $(2,0)$ $\mathbb{Z}_l$}}_{\textrm{twist}}[\mathbb{R}^5;N](x_1;x_2)}
{\mathcal{I}^{\textrm{6d $(2,0)$ $\mathbb{Z}_l$}}_{\textrm{twist}}[\mathbb{R}^5;N=\infty](x_1;x_2)}
&=\sum_{m=0}^{\infty}
x_1^{lNm}\hat{G}_{m}^{\mathbb{C}^2/\mathbb{Z}_l}(x_1;x_2). 
\end{align}
We find that 
\begin{align}
\hat{G}_{m}^{\mathbb{C}^2/\mathbb{Z}_l}(x_1;x_2)
&=\sigma_1 \Bigl[ \mathcal{I}^{\textrm{$U(m)$ ADHM-[$l$]}}_C[\mathbb{C}^2\times \mathbb{C}^2/\mathbb{Z}_l;m](x_1;x_2) \Bigr]. 
\end{align}
In other words, the index of the M5-giants of wrapping number $lm$ in the $S^5/\mathbb{Z}_l$ is beautifully given by 
the Coulomb index of the $U(m)$ ADHM theory under the transformation (\ref{var_change})! 

The giant graviton expansion (\ref{GG6d_lC}) is well-defined in the unrefined limit 
where the expansion coefficients are given by the Coulomb index (\ref{ind_ADHMunlC_C^2/Z_l}) 
with the specialization $x_2$ $\rightarrow$ $x_1^{-1}$. 
Also, the expansion holds in the half-BPS limit. 
It follows that the half-BPS index for the M2-brane SCFT is equal to that for the M5-brane SCFT. 
We find 
\begin{align}
\label{1/2M2_Zl}
\mathcal{I}^{\textrm{$U(N)$ ADHM-[$l$]}}_{\textrm{$\frac12$BPS},C}[\mathbb{C}^2\times \mathbb{C}^2/\mathbb{Z}_{l};N](x_1)
&=\prod_{n=1}^{N}\frac{1}{1-x_1^{ln}}, \\
\mathcal{I}^{\textrm{6d $(2,0)$ $\mathbb{Z}_l$}}_{\textrm{$\frac12$BPS}}[\mathbb{R}^5;N](x_1)
&=\prod_{n=1}^{N}\frac{1}{1-x_1^{ln}}. 
\end{align}
It can be shown that they have the giant graviton expansions of the form
\begin{align}
\frac{\mathcal{I}^{\textrm{$U(N)$ ADHM-[$l$]}}_{\textrm{$\frac12$BPS},C}[\mathbb{C}^2\times \mathbb{C}^2/\mathbb{Z}_l;N](x_1)}
{\mathcal{I}^{\textrm{$U(\infty)$ ADHM-[$l$]}}_{\textrm{$\frac12$BPS},C}[\mathbb{C}^2\times \mathbb{C}^2/\mathbb{Z}_l;N=\infty](x_1)}
&=\sum_{m=0}^{\infty}x_1^{lNm}\hat{f}_{m}^{\mathbb{C}^2/\mathbb{Z}_l}(x_1), \\
\frac{\mathcal{I}^{\textrm{6d $(2,0)$ $\mathbb{Z}_l$}}_{\textrm{$\frac12$BPS}}[\mathbb{R}^5;N](x_1)}
{\mathcal{I}^{\textrm{6d $(2,0)$ $\mathbb{Z}_l$}}_{\textrm{$\frac12$BPS}}[\mathbb{R}^5;N=\infty](x_1)}
&=\sum_{m=0}^{\infty}x_1^{lNm}\hat{g}_{m}^{\mathbb{C}^2/\mathbb{Z}_l}(x_1), 
\end{align}
where 
\begin{align}
\hat{f}_{m}^{\mathbb{C}^2/\mathbb{Z}_l}(x_1)&=\frac{(-1)^m x_1^{\frac{l m(m+1)}{2}}}{(x_1^l;x_1^l)_{m}}
=\mathcal{I}^{\textrm{6d $(2,0)$ $\mathbb{Z}_l$}}_{\textrm{$\frac12$BPS}}[\mathbb{R}^5;m](x_1^{-1})
\end{align}
and
\begin{align}
\hat{g}_{m}^{\mathbb{C}^2/\mathbb{Z}_l}(x_1)&=\frac{(-1)^m x_1^{\frac{l m(m+1)}{2}}}{(x_1^l;x_1^l)_{m}}
=\mathcal{I}^{\textrm{$U(m)$ ADHM-}[l]}_{\textrm{$\frac12$BPS},C}[\mathbb{C}^2\times \mathbb{C}^2/\mathbb{Z}_l;m](x_1^{-1})
\end{align}
are the index for the half-BPS M5-brane giants of mod $l$ wrapping number $lm$ in $S^7/\mathbb{Z}_l$ 
and that for the half-BPS M2-brane giants of mod $l$ wrapping number $lm$ in $S^4/\mathbb{Z}_l$ respectively. 

In addition, the indices have the double sum giant graviton expansions
\begin{align}
\frac{\mathcal{I}^{\textrm{$U(N)$ ADHM-[$l$]}}_{C}[\mathbb{C}^2\times \mathbb{C}^2/\mathbb{Z}_l;N](x_1;x_2)}
{\mathcal{I}^{\textrm{$U(\infty)$ ADHM-[$l$]}}_{C}[\mathbb{C}^2\times \mathbb{C}^2/\mathbb{Z}_l;N=\infty](x_1;x_2)}
&=\sum_{m_1=0}^{\infty}
\sum_{m_2=0}^{\infty}
x_1^{lNm_1}
x_2^{lNm_2}
(x_1 x_2)^{m_1 m_2}
\nonumber\\
&\times 
\hat{F}_{m_1}^{\mathbb{C}^2/\mathbb{Z}_l}(x_1;x_2)
\hat{F}_{m_2}^{\mathbb{C}^2/\mathbb{Z}_l}(x_2;x_1) 
\end{align}
and 
\begin{align}
\frac{\mathcal{I}^{\textrm{6d $(2,0)$ $\mathbb{Z}_l$}}_{\textrm{twist}}[\mathbb{R}^5;N](x_1;x_2)}
{\mathcal{I}^{\textrm{6d $(2,0)$ $\mathbb{Z}_l$}}_{\textrm{twist}}[\mathbb{R}^5;N=\infty](x_1;x_2)}
&=\sum_{m_1=0}^{\infty}
\sum_{m_2=0}^{\infty}
x_1^{lNm_1}
x_2^{lNm_2}
(x_1 x_2)^{m_1 m_2}
\nonumber\\
&\times 
\hat{G}_{m_1}^{\mathbb{C}^2/\mathbb{Z}_l}(x_1;x_2)
\hat{G}_{m_2}^{\mathbb{C}^2/\mathbb{Z}_l}(x_2;x_1). 
\end{align}

\subsubsection{$\mathbb{C}^2/\widehat{D}_1$}
When a stack of $N$ M2-branes probes the $D$-type singularity, 
one has the effective theory on the M2-branes with orthogonal or symplectic gauge group. 
For $USp(2N)$ ADHM theory with a hypermultiplet in the antisymmetric representation as well as $6$ flavors, 
one has a dual description as the $U(N)$ ADHM theory with $4$ flavors \cite{deBoer:1996mp}. 
We have \cite{Hayashi:2022ldo}
\begin{align}
\mathcal{I}^{\textrm{$USp(2N)+$asym-[6]}}_{C}[\mathbb{C}^2\times \mathbb{C}^2/\widehat{D}_1;N](x_1;x_2)
&=\mathcal{I}^{\textrm{$U(N)$ ADHM-[4]}}_{C}[\mathbb{C}^2\times \mathbb{C}^2/\mathbb{Z}_4;N](x_1;x_2). 
\end{align}
Therefore, it also admits the giant graviton expansions. 
However, unlike the theories with unitary gauge group, the Coulomb indices do not contain the topological fugacity in general. 

\subsection{Higgs indices}
\label{sec_Higgs}
Now we present the giant graviton expansions of the Higgs indices. 

\subsubsection{$\mathcal{N}=8$ $U(N)$ SYM theory}
\label{sec_N8SYMHiggs}
Our starting point of the Higgs indices is $\mathcal{N}=8$ $U(N)$ SYM theory. 
The theory is a world-volume theory on a stack of $N$ D2-branes in Type IIA string theory \cite{Witten:1995im}. 
The classical moduli space of vacua is $(\mathbb{R}^{7}\times S^1)^N/S_N$ 
where $\mathbb{R}^7$ is parameterized by seven adjoint scalar fields corresponding to the $7$ transverse directions of the D2-branes, 
$S^1$ is parameterized by dual photons corresponding to the M-theory circle 
and $S_N$ is the symmetric group. 
As the radius of the torus of the moduli space is proportional to the coupling constant $g^2$,  
the moduli space becomes \cite{Seiberg:1997ax}
\begin{align}
\mathcal{M}_A&=\mathrm{Sym}^N(\mathbb{C}^4)
\end{align}
in the strong coupling limit. 
The factor $\mathbb{C}^4$ corresponds to the spacetime transverse to the lifted M2-branes in M-theory. 

While the supersymmetric indices for $\mathcal{N}=8$ SYM theory are not simply computed from the localization formula 
as the associated infinite sums over the magnetic fluxes are not convergent, 
one can get the well-defined Higgs indices which enumerate the gauge invariant BPS local operators consisting of an adjoint hypermultiplet. 
The Higgs index of $\mathcal{N}=8$ $U(N)$ SYM theory is given by \cite{Hayashi:2024jof}\footnote{
We note that the expression coincides with 
the large $N$ normalized two-point function of the (anti)symmetric Wilson line operators 
in $\mathcal{N}=4$ $U(N)$ SYM theory \cite{Hatsuda:2023iwi,Hatsuda:2023imp}. 
}
\begin{align}
\label{ind_N8A_H}
\mathcal{I}^{\textrm{$\mathcal{N}=8$ $U(N)$}}_{H}(x_1;x_2)
&=\sum_{k=0}^{N}\frac{x_1^k}{(x_1;x_1)_{k}(x_2;x_2)_{N-k}}. 
\end{align}
In the large $N$ limit it becomes
\begin{align}
\mathcal{I}^{\textrm{$\mathcal{N}=8$ $U(\infty)$}}_{H}(x_1;x_2)
&=\prod_{n=1}^{\infty}\frac{1}{(1-x_1^n)(1-x_2^n)}. 
\label{N8ANinfty}
\end{align}
It admits the giant graviton expansion of the form 
\begin{align}
\frac{\mathcal{I}^{\textrm{$\mathcal{N}=8$ $U(N)$}}_{H}(x_1;x_2)}
{\mathcal{I}^{\textrm{$\mathcal{N}=8$ $U(\infty)$}}_{H}(x_1;x_2)}
&=\sum_{m=0}^{\infty}x_1^{Nm} \hat{\mathcal{F}}^{(A)_m}_{H}(x_1;x_2), 
\label{N8Asinglesumgg}
\end{align}
where
\begin{align}
\hat{\mathcal{F}}^{(A)_m}_{H}(x_1;x_2)
&=(x_2;x_2)_{\infty}
\frac{(-1)^{m-1}x_1^{\frac{m(m-1)}{2}}}
{(x_1;x_1)_{m-1}}\sum_{l=0}^{\infty}
\frac{x_1^{-ml}}{(x_2;x_2)_{l}}. 
\end{align}

Under the changes (\ref{var_change}) of variables we find 
\begin{align}
\label{ind_A}
\mathcal{F}^{(A)_N}_{H}(x_1;x_2)&=\sigma_1 
\Bigl[\hat{\mathcal{F}}^{(A)_N}_{H}(x_1;x_2)\Bigr]
\nonumber\\
&=\frac{(x_1x_2;x_1x_2)_{\infty}}
{(x_1;x_1)_{N-1}(x_1^{N};x_1x_2)_{\infty}}. 
\end{align}
It would be tempting to explore any other approach to derive the dual index (\ref{ind_A}). 
Taking into account the fact that 
the giant graviton M5-branes wrapping $S^5$ can reduce to D4-branes wrapping $\mathbb{CP}^2$ 
after the reduction of M-theory to Type IIA string theory, 
it may be natural to expect that 
the index (\ref{ind_A}) has an interpretation in terms of a certain supersymmetric partition function of 5d SYM theory on $S^1\times \mathbb{CP}^2$. 
We leave it to an interesting future work. 

Conversely, let us consider the giant graviton expansion of the dual index (\ref{ind_A})
\begin{align}
\frac{\mathcal{F}^{(A)_{N}}_{H}(x_1;x_2)}
{\mathcal{F}^{(A)_{\infty}}_{H}(x_1;x_2)}
&=\sum_{m=0}^{\infty}
x_1^{Nm}\hat{\mathcal{G}}_{H}^{(A)_{m}}(x_1;x_2). 
\end{align}
We find that the giant graviton index which encodes the finite $N$ correction is again given 
by the Higgs index  (\ref{ind_N8A_H}) for $\mathcal{N}=8$ $U(m)$ SYM theory
\begin{align}
\hat{\mathcal{G}}_{H}^{(A)_{m}}(x_1;x_2)
&=\sigma_1\Bigl[
\mathcal{I}^{\textrm{$\mathcal{N}=8$ $U(m)$}}_H(x_1;x_2)
\Bigr]. 
\end{align}

When we take the half-BPS limit, the Higgs index (\ref{ind_N8A_H}) becomes
\begin{align}
\mathcal{I}^{\textrm{$\mathcal{N}=8$ $U(N)$}}_{\textrm{$\frac12$BPS}}(x_1)
&=\prod_{n=1}^N \frac{1}{1-x_1^n}
\end{align}
and the dual index (\ref{ind_A}) becomes 
\begin{align}
\label{ind_A_1/2BPS}
\mathcal{F}^{(A)_N}_{\textrm{$\frac12$BPS,$H$}}(x_1)&=\frac{1}{(x_1;x_1)_N}. 
\end{align}
Since they coincide with the half-BPS index (\ref{1/2M2}) of $N$ M2-branes and the half-BPS index (\ref{1/2M5}) of $N$ M5-branes respectively, 
they admit the giant graviton expansions with the same structures as (\ref{1/2M2gg}) and (\ref{1/2M5gg}). 

Also we have the double sum giant graviton expansions 
\begin{align}
\frac{\mathcal{I}^{\textrm{$\mathcal{N}=8$ $U(N)$}}_{H}(x_1;x_2)}
{\mathcal{I}^{\textrm{$\mathcal{N}=8$ $U(\infty)$}}_{H}(x_1;x_2)}
&=\sum_{m_1=0}^{\infty}
\sum_{m_2=0}^{\infty}
x_1^{Nm_1}x_2^{Nm_2}
(x_1x_2)^{m_1 m_2}
\nonumber\\
&\times 
\hat{\mathcal{F}}^{(A)_{m_1}}_{H}(x_1;x_2)
\hat{\mathcal{F}}^{(A)_{m_2}}_{H}(x_2;x_1), 
\end{align}
and 
\begin{align}
\frac{\mathcal{F}^{(A)_N}_{H}(x_1;x_2)}
{\mathcal{F}^{(A)_{\infty}}_{H}(x_1;x_2)}
&=\sum_{m_1=0}^{\infty}
\sum_{m_2=0}^{\infty}
x_1^{Nm_1}
x_2^{Nm_2}
(x_1 x_2)^{m_1m_2}
\nonumber\\
&\times 
\hat{\mathcal{G}}^{(A)}_{m_1}(x_1;x_2)
\hat{\mathcal{G}}^{(A)}_{m_2}(x_2;x_1). 
\end{align}

\subsubsection{$\mathcal{N}=8$ $USp(2N)$, $SO(2N+1)_{+}$ and $O(2N)_{+}$ SYM theories}
\label{sec_N8BCDSYMHiggs}
Next consider $\mathcal{N}=8$ SYM theory with other gauge groups. 
We note that $\mathcal{N}=8$ $B_N=SO(2N+1)$, $C_N=USp(2N)$ and $D_N=O(2N)_+$ SYM theories\footnote{
Here the subscript of $SO(2N+1)$ gauge group stands for the discrete fugacity $\chi=+$ for the $\mathbb{Z}_2$ charge conjugation symmetry $\mathbb{Z}^{\mathcal{C}}_{2}$, while the subscript of $O(2N)$ gauge group indicate the discrete fugacity $\chi'=+$ for which the gauge invariant minimal monopole operator is charge conjugation even \cite{Aharony:2013kma}.
}
have the same moduli space in the strong coupling limit \cite{Berkooz:1998sn}
\begin{align}\label{MBCD}
\mathcal{M}_{BCD}&=\mathrm{Sym}^N (\mathbb{C}^4/\mathbb{Z}_2). 
\end{align}
Accordingly, the Higgs indices of $SO(2N+1)$ theory with $\chi=+$, 
$USp(2N)$ theory and $O(2N)$ theory with $\chi'=+$ agree with each other
\begin{align}
\label{ind_N8BCD_H}
\mathcal{I}^{\textrm{$\mathcal{N}=8$ $BCD_{N}$}}_{H}(x_1;x_2)
&=\mathcal{I}^{\textrm{$\mathcal{N}=8$ $USp(2N)$}}_{H}(x_1;x_2)
\nonumber\\
&=\mathcal{I}^{\textrm{$\mathcal{N}=8$ $SO(2N+1)_{ +}$}}_{H}(x_1;x_2)
\nonumber\\
&=\mathcal{I}^{\textrm{$\mathcal{N}=8$ $O(2N)_{+}$}}_{H}(x_1;x_2). 
\end{align}
The $SO(2N+1)$, $USp(2N)$ or $O(2N)$ theory appears as an effective theory of a stack of $N$ D2-branes 
in the background of $\widetilde{\textrm{O2}^-}$, O2$^+$ or O2$^-$. 
The factor $\mathbb{C}^4/\mathbb{Z}_2$ in the moduli space (\ref{MBCD}) corresponds to the $\mathbb{Z}_2$ orbifold geometry 
transverse to the lifted M2-branes.

To proceed, let us introduce the grand canonical Higgs index defined by
\begin{align}
\label{grand_BCD}
\Xi^{(BCD)}(\kappa;x_1;x_2)&=\sum_{N=0}^{\infty}
\mathcal{I}^{\textrm{$\mathcal{N}=8$ $BCD_N$}}_{H}(x_1;x_2)\kappa^N
\nonumber\\
&=\exp\left[
\sum_{m=1}^{\infty}
\frac{\mathcal{I}^{\textrm{$\mathcal{N}=8$ $BCD_1$}}_{H}(x_1^m;x_2^m)}{m}
\kappa^m
\right], 
\end{align}
where 
\begin{align}
\mathcal{I}^{\textrm{$\mathcal{N}=8$ $BCD_1$}}_{H}(x_1;x_2)
&=\frac{1+x_1x_2-x_1 x_2(x_1+x_2)}
{(1-x_1^2) (1-x_2^2)}
\end{align}
is the Higgs index of $\mathcal{N}=8$ $USp(2)$, $SO(3)_{+}$ or $O(2)_{+}$ SYM theory. 
The grand canonical index (\ref{grand_BCD}) can be evaluated as
\begin{align}
\Xi^{(BCD)}(\kappa;x_1;x_2)&=
\prod_{n=0}^{\infty}\prod_{m=0}^{\infty}
\frac{(1-x_1^{2n+2}x_2^{2m+1}\kappa) (1-x_1^{2n+1}x_2^{2m+2}\kappa)}
{(1-x_1^{2n} x_2^{2m}\kappa) (1-x_1^{2n+1} x_2^{2m+1}\kappa)}. 
\label{gcN8BCD}
\end{align}

We find that the Higgs index has the single sum giant graviton expansion
\begin{align}
\frac{\mathcal{I}^{\textrm{$\mathcal{N}=8$ $BCD_N$}}_{H}(x_1;x_2)}
{\mathcal{I}^{\textrm{$\mathcal{N}=8$ $BCD_{\infty}$}}_{H}(x_1;x_2)}
&=\sum_{m=0}^{\infty}x_1^{2Nm} \hat{\mathcal{F}}^{(BCD)_m}_{H}(x_1;x_2), 
\label{N8BCDsinglesumgg}
\end{align}
where
\begin{align}\label{largeNIBCD}
&{\cal I}^{{\cal N}=8\,BCD_\infty}_H(x_1;x_2)=\prod_{\substack{b,c\ge 0\\ (b,c)\neq (0,0)}}\frac{1}{1-x_1^{2b}x_2^{2c}}
\prod_{b,c=0}^\infty\frac{(1-x_1^{2b+2}x_2^{2c+1})(1-x_1^{2b+1}x_2^{2c+2})}{1-x_1^{2b+1}x_2^{2c+1}}
\end{align}
and
\begin{align}
\hat{\mathcal{F}}^{(BCD)_m}_{H}(x_1;x_2)
&=\prod_{k=1}^{m}
\frac{(x_1^{-2k+1}x_2^{2};x_2^2)_{\infty} (x_1^{-2k+2}x_2;x_2^2)_{\infty}}
{(x_1^{-2k}; x_2^2)_{\infty} (x_1^{-2k+1}x_2;x_2^2)_{\infty}}. 
\end{align}
Again the expansion coefficients in the giant graviton expansion (\ref{N8BCDsinglesumgg}) 
will be interpreted as the BPS indices of D4-brane giant gravitons. 
We see that the D4-brane giant graviton carries the $\mathbb{Z}_2$ charges. 
This is consistent with the fact that 
it is lifted to M-theory on the $\mathbb{Z}_2$ orbifold geometry 
which can support the orbifold M5-brane giant gravitons wrapping a non-trivial $\mathbb{Z}_2$ homology cycle. 

The dual index is
\begin{align}
\label{ind_BCD}
\mathcal{F}^{(BCD)_N}_{H}(x_1;x_2)&=\sigma_1 
\Bigl[\hat{\mathcal{F}}^{(BCD)_N}_{H}(x_1;x_2)\Bigr]
\nonumber\\
&=\prod_{k=1}^{N}
\frac{(x_1^{2k+1}x_2^2;x_1^2 x_2^2)_{\infty} (x_1^{2k-1}x_2;x_1^2x_2^2)_{\infty}}
{(x_1^{2k};x_1^2 x_2^2)_{\infty} (x_1^{2k}x_2;x_1^2x_2^2)_{\infty}}. 
\end{align}

We find that the dual index also obeys the inverse relation
\begin{align}
\frac{\mathcal{F}^{(BCD)_N}_{H}(x_1;x_2)}
{\mathcal{F}^{(BCD)_{\infty}}_{H}(x_1;x_2)}
&=\sum_{m=0}^{\infty}
x_1^{2Nm}\hat{\mathcal{G}}^{(BCD)_m}_{H}(x_1;x_2),
\end{align}
where 
\begin{align}
\hat{\mathcal{G}}^{(BCD)_m}_{H}(x_1;x_2)
&=\sigma_1\Bigl[
\mathcal{I}^{\textrm{$\mathcal{N}=8$ $BCD_m$}}_{H}(x_1;x_2)
\Bigr]. 
\end{align}

In the half-BPS limit, the Higgs index (\ref{ind_N8BCD_H}) and the dual index (\ref{ind_BCD}) reduce to
\begin{align}
\mathcal{I}^{\textrm{$\mathcal{N}=8$ $BCD_N$}}_{\textrm{$\frac12$BPS}}(x_1)
&=\prod_{n=1}^N \frac{1}{1-x_1^{2n}}, \\
\mathcal{F}^{(BCD)_N}_{\textrm{$\frac12$BPS}}(x_1)&=\prod_{n=1}^N \frac{1}{1-x_1^{2n}}. 
\end{align}
They agree with the half-BPS index (\ref{1/2M2Z2}) for $N$ M2-branes probing the orbifold geometry involving $\mathbb{C}^2/\mathbb{Z}_2$ 
and the half-BPS index (\ref{1/2M5Z2}) of $N$ M5-branes on the orbifold space $S^5/\mathbb{Z}_2$ respectively. 
Thus they have the giant graviton expansions with the same structures as (\ref{1/2M2Z2gg}) and (\ref{1/2M5Z2gg}). 

Also we find that it has the double sum giant graviton expansions
\begin{align}
\frac{\mathcal{I}^{\textrm{$\mathcal{N}=8$ $BCD_N$}}_{H}(x_1;x_2)}
{\mathcal{I}^{\textrm{$\mathcal{N}=8$ $BCD_{\infty}$}}_{H}(x_1;x_2)}
&=\sum_{m_1=0}^{\infty}
\sum_{m_2=0}^{\infty}
x_1^{2Nm_1}x_2^{2Nm_2}
(x_1x_2)^{m_1 m_2}
\nonumber\\
&\times 
\hat{\mathcal{F}}^{(BCD)_{m_1}}_{H}(x_1;x_2)
\hat{\mathcal{F}}^{(BCD)_{m_2}}_{H}(x_2;x_1). 
\end{align}
and 
\begin{align}
\frac{\mathcal{F}^{(BCD)_N}_{H}(x_1;x_2)}
{\mathcal{F}^{(BCD)_{\infty}}_{H}(x_1;x_2)}
&=\sum_{m_1=0}^{\infty}
\sum_{m_2=0}^{\infty}
x_1^{2Nm_1}
x_2^{2Nm_2}
(x_1 x_2)^{m_1m_2}
\nonumber\\
&\times 
\hat{\mathcal{G}}^{(BCD)_{m_1}}_{H}(x_1;x_2)
\hat{\mathcal{G}}^{(BCD)_{m_2}}_{H}(x_2;x_1). 
\end{align}

\subsubsection{${\cal N}=8$ $SO(2N)$ SYM theory}
We consider giant graviton expansions of the Higgs index of the 3d $\mathcal{N}=8$ $SO(2N)$ SYM theory. 
The $SO(2N)$ theory can be viewed as a ``parent theory'' of $O(2N)$ theory 
since $O(2N)$ theory is obtained by gauging the $\mathbb{Z}_2$ charge conjugation symmetry $\mathbb{Z}_{\mathcal{C}}$ of $SO(2N)$ theory. 
Holographically, the ungauging procedure of the discrete $\mathbb{Z}_2$ symmetry in the $O(2N)$ gauge theory can be obtained by taking 
the vanishing Dirichlet boundary condition for the 1-form $\mathbb{Z}_2$ gauge field in the supergravity background \cite{Aharony:2016kai}. 
One remarkable feature of the $SO(2N)$ theory 
is that it has the Pfaffian operator, a gauge invariant chiral primary operator consisting of $N$ scalar fields. 
The number of Pfaffian operators is conserved modulo 2 so that 
the moduli space of the 3d $\mathcal{N}=8$ $SO(2N)$ SYM theory is given by
\be
\mathcal{M}_{SO(2N)} = \text{Sym}^N\left(\mathbb{C}^4/\mathbb{Z}_2\right) \times \mathbb{Z}_2,
\ee
which is different from \eqref{MBCD}. 
Accordingly, the index of the 3d $\mathcal{N}=8$ $SO(2N)$ SYM theory depends on an additional discrete fugacity $\chi$ for $\mathbb{Z}_{\mathcal{C}}$ 
and therefore it is different from that of the 3d $\mathcal{N}=8$ $O(2N)$ SYM theory considered in section \ref{sec_N8BCDSYMHiggs}. 
For $\chi=+$ the index should have contributions from the Pfaffian operators. 

The grand canonical Higgs index of the 3d $\mathcal{N}=8$ $SO(2N)$ SYM theory with $\chi=+$ is defined by
\be\label{grand_SO}
\Xi^{SO_+}(\kappa; x_1; x_2) = 1 + \sum_{N=1}^{\infty}\mathcal{I}_H^{\mathcal{N}=8\;SO(2N)_{+}}(x_1; x_2)\kappa^N, 
\ee
where $\mathcal{I}_H^{\mathcal{N}=8\;SO(2N)_+}(x_1; x_2)$ is the Higgs index of the 3d $\mathcal{N}=8$ $SO(2N)$ SYM theory with $\chi=+$. 
We propose that the grand canonical Higgs index \eqref{grand_SO} is given by 
\be\label{gcN8SO}\begin{split}
&\Xi^{SO_+}(\kappa;x_1;x_2) = \prod_{m=0}^{\infty}\prod_{n=0}^{\infty}\frac{\left(1-\kappa x_1^{2m+2}x_2^{2n+1}\right)\left(1-\kappa x_1^{2m+1}x_2^{2n+2}\right)}{\left(1-\kappa x_1^{2m}x_2^{2n}\right)\left(1 - \kappa x_1^{2m+1}x_2^{2n+1}\right)}\cr
&\hspace{3cm}  + \prod_{m=0}^{\infty}\prod_{n=0}^{\infty}\frac{\left(1-\kappa x_1^{2m+1}x_2^{2n+1}\right)\left(1-\kappa x_1^{2m+2}x_2^{2n+2}\right)}{\left(1-\kappa x_1^{2m}x_2^{2n+1}\right)\left(1 - \kappa x_1^{2m+1}x_2^{2n}\right)}  - 1.
\end{split}\ee
The index $\mathcal{I}_H^{\mathcal{N}=8\;SO(N)}(x_1; x_2)$ in the limit $N \to \infty$ is the same as the one in \eqref{largeNIBCD}.  
We find that the Higgs index has the single sum giant graviton expansion
\begin{align}
\frac{
{\cal I}^{{\cal N}=8\;SO(2N)_+}_H(x_1;x_2)
}{
{\cal I}^{{\cal N}=8\;SO(\infty)_+}_H(x_1;x_2)
}
&=\sum_{m=0}^{\infty}x_1^{2mN}
\left(
1+x_1^{N} \frac{(x_1^{-2m}x_2;x_2^2)_{\infty}}{(x_1^{-2m-1}x_2;x_2^2)_{\infty}}
\right)
\hat{\mathcal{F}}^{(BCD)_{m}}_{H}(x_1;x_2), 
\label{SO2Nsinglesumgg}
\end{align}
where the second term in the curly bracket 
is the expected contributions from the protected sector with the Pfaffian operators. 

In the half-BPS limit the Higgs index becomes 
\begin{align}
\mathcal{I}^{{\mathcal{N}}=8\;SO(2N)_+}_{\textrm{$\frac12$BPS}}(x_1)
&=\frac{1}{1-x_1^N}\prod_{n=1}^{N-1}\frac{1}{1-x_1^{2n}}. 
\end{align}
One can show that it has the following giant graviton expansion
\begin{align}
\frac{\mathcal{I}^{{\mathcal{N}}=8\;SO(2N)_+}_{\textrm{$\frac12$BPS}}(x_1)}
{\mathcal{I}^{{\mathcal{N}}=8\;SO(\infty)_+}_{\textrm{$\frac12$BPS}}(x_1)}
&=
(1+x_1^N)\sum_{m=0}^{\infty}
x_1^{2mN} \hat{\mathcal{F}}^{(BCD)_m}_{\textrm{$\frac12$BPS}}(x_1). 
\end{align}
Here the second term in the curly bracket 
corresponds to the protected sector containing the Pfaffian operator. 


\subsubsection{$U(N)$ ADHM theories}
Let us consider the supersymmetric index for the $U(N)$ ADHM theory with $l$ flavors in the Higgs limit \eqref{def_Hindex}. 
For $l=0$ and $l=1$, this setup reduces respectively to the ${\cal N}=8$ $U(N)$ SYM theory considered in section \ref{sec_N8SYMHiggs} 
and the $U(N)$ ADHM theory with one flavor in the Coulomb limit considered in section \ref{sec_CoulombC2}.
For $l\ge 2$ it is more difficult to derive the giant graviton expansions since the closed-form expressions for the Higgs indices are not known for general $N$. 
Nevertheless it is possible to generate the Higgs indices for first a few $N$'s either in the closed form or in the small $\mathfrak{t}$ expansion to a sufficiently high order
by using the relation between the Higgs index and the instanton partition function of 5d ${\cal N}=1$ $SU(l)$ Yang-Mills theory with instanton number $N$ \eqref{Nekrasov}.
By using such data we find the leading coefficient of the giant graviton expansion as
\begin{align}
\label{GGexp_ADHM_H}
\frac{
{\cal I}^{U(N)\text{ADHM-}[l]}_H[\mathbb{C}^2\times \mathbb{C}^2/\mathbb{Z}_l;N](x_1;x_2)
}{
{\cal I}^{U(\infty)\text{ADHM-}[l]}_H[\mathbb{C}^2\times \mathbb{C}^2/\mathbb{Z}_l;N=\infty](x_1;x_2)
}=1+
x_1^{N}{\hat {\cal F}}_1^{(l)}(x_1;x_2)
+{\cal O}(x_1^{2N}),
\end{align}
where we have set the fugacities $y_\alpha$ for the flavor symmetry to $y_\alpha=1$.
Here the leading giant graviton coefficient ${\cal F}^{(l)}_1(x_1;x_2)$ is given by\footnote{See appendix \ref{app_UNlHgg} for the detail.}
\begin{align}
\label{GGind_ADHM_H}
{\hat {\cal F}}^{(l)}_{1}(x_1;x_2)
&=\frac{1}{(x_1^{-1};x_2)_{\infty}}
\chi_{\widehat{\mathfrak{su}}(l)_{1}}(x_2),
\end{align}
where 
\begin{align}
\label{sulWZW_ch}
\chi_{\widehat{\mathfrak{su}}(l)_{1}}(q)
&=
\frac{1}{(q;q)_{\infty}^{l-1}}
\sum_{
\begin{smallmatrix}
m_1,\cdots, m_l\in \mathbb{Z}\\
m_1+m_2+\cdots+m_l=0\\
\end{smallmatrix}
}q^{\sum_{i=1}^{l}\frac{m_i^2}{2}}
\end{align}
is the vacuum character of the $SU(l)_{1}$ WZW model of level $1$. 
As the WZW model admits the coset construction 
\begin{align}
SU(l)_{1}&=\frac{U(l)_1}{U(1)_{-l}}, 
\end{align}
the vacuum character (\ref{sulWZW_ch}) can be also obtained from a system of $l$ free fermions coupled to the $U(1)$ Chern-Simons theory with level $-l$
\begin{align}
\chi_{\widehat{\mathfrak{su}}(l)_{1}}(q)&=
(q;q)_{\infty}
\oint \frac{ds}{2\pi is}
(s;q)_{\infty}^l (qs^{-1};q)_{\infty}^l. 
\end{align}
This is nothing but a level-rank duality between $SU(l)_{1}$ and $U(1)_{-l}$. 
After multiplying the character (\ref{sulWZW_ch}) by the generation function for the ordinary partitions, i.e. $1/(q;q)_{\infty}$, 
it can be also viewed as a generating function for the generalized Frobenius partitions $c\phi_l(n)$ of $n$ with $l$ colors \cite{MR743546}. 

Note that when $l=1$, the leading term (\ref{GGind_ADHM_H}) in the giant graviton expansion (\ref{GGexp_ADHM_H}) has no contribution from the WZW character (\ref{sulWZW_ch})
so that it reduces to the index (\ref{GGind_ADHMu11C}) for a single M5-brane giant graviton wrapped on the $5$-cycle in $S^7$. 
Thus the appearance of the WZW model stems from the $A_{l-1}$ singularity due to the presence of $l>1$ KK monopoles \cite{Townsend:1995kk} in M-theory. 
In fact, a similar configuration of M5-branes described by the WZW model or the free fermion systems 
can be found in the literature \cite{Itzhaki:2005tu,Dijkgraaf:2007sw,Dijkgraaf:2007fe,Tan:2008wp,Witten:2009at,Ohlsson:2012yn,Lambert:2018mfb,Gustavsson:2022jpo}. 
The transverse space to the $l$ KK monopoles is described by an $l$-centered multi-centered Taub-NUT space $TN_l$ 
that admits a triholomorphic $U(1)$ symmetry preserving the hyper-K\"{a}hler structure. 
The $U(1)$ action acts on the $TN_l$ except for two-dimensional fixed point locus $\Sigma$. 
When a system of M5-branes is wrapped on $TN_l$, 
the anomalous gauge transformation of the topological term induces the $SU(l)$ WZW model of level $1$ on $\Sigma$ \cite{Witten:2009at}. 
The M-theory configuration is equivalent to Type IIA string theory involving D4-branes intersecting with $l$ D6-branes on the two-dimensional surface $\Sigma$, 
which is referred to as the I-brane \cite{Itzhaki:2005tu}. 
The massless modes of D4-D6 strings are chiral fermions supported on the I-brane 
which realize the WZW model \cite{Dijkgraaf:2007sw}. 
It would be interesting to extend the analysis to the cases with more than a single M5-brane giant graviton.

Besides, we can obtain the exact form for the Higgs indices in the half-BPS limit
by making use of the orthogonality of the Hall-Littlewood functions \cite{MR1354144} in the integral formula.
One finds that
\begin{align}
\label{1/2M2_Zl_H}
\mathcal{I}^{\textrm{$U(N)$ ADHM-[$l$]}}_{\textrm{$\frac12$BPS, $H$}}[\mathbb{C}^2\times \mathbb{C}^2/\mathbb{Z}_{l};N](x_1)
&=\prod_{n=1}^{N}\frac{1}{1-x_1^n}. 
\end{align}
As the theories are not self-mirror for $l\ge2$, the expression (\ref{1/2M2_Zl_H}) is not equal to the half-BPS limit (\ref{1/2M2_Zl}) of the Coulomb index for the same theory, 
but rather coincide with the half-BPS limit (\ref{1/2M2}) of the Coulomb index or equivalently the Higgs index of the $U(N)$ ADHM theory with one flavor. 
In other words, the half-BPS limits of the Higgs indices of the ADHM theories are independent of $l$. 
Therefore the indices have the giant graviton expansions of the form (\ref{1/2M2gg}), 
which indicates the contributions from the spherical half-BPS giant gravitons of wrapping numbers being arbitrary non-negative integers.

\section*{Acknowledgement}
The authors would like to thank Hai Lin, Satoshi Nawata and Shingo Takeuchi for useful discussions and comments. 
The work of H.H.~is supported in part by JSPS KAKENHI Grand Number JP23K03396. 
The work of T.O.~was supported by the Startup Funding no.~4007012317 of the Southeast University. 
The work of T.N.~was supported by the Startup Funding no.~2302-SRFP-2024-0012 of Shanghai Institute for Mathematics and Interdisciplinary Sciences.
Preliminary results of this paper were presented in the international conferences ``Satellite conference of 2024 International Congress of Basic Science in Mathematical Physics'' held at Fudan University, Shanghai, China and ``7TH INTERNATIONAL CONFERENCE ON HOLOGRAPHY AND STRING THEORY IN DA NANG'' held at Duy Tan University, Da Nang, Vietnam.
Part of the results was computed by using the high performance computing facility provided by Yukawa Institute for Theoretical Physics (Sushiki server).

\appendix
\section{Grand canonical indices and expansion formulas}
\label{app_ggfromgc}
In this appendix we present several remarks on a useful approach to find giant graviton expansions using the grand canonical indices.

\subsection{Coulomb indices}
Let us start with the Coulomb index of $U(N)$ ADHM theory with $l$ flavors, whose grand canonical index is given as \eqref{gcUNlADHMC}. 
We can formally invert the relation to write the Coulomb index as
\begin{align}
{\cal I}^{U(N)\text{ ADHM-}[l]}_C[\mathbb{C}^2\times \mathbb{C}^2/\mathbb{Z}_l;N](x_1;x_2)=
\oint \frac{d\kappa}{2\pi i\kappa}\kappa^{-N} \Xi^{\mathbb{C}^2/\mathbb{Z}_l}_{C}(\kappa;x_1;x_2).
\label{inversetrsf}
\end{align}
Here the integration contour is a circle around the origin which is sufficiently small so that all the poles of 
the grand canonical index at $\kappa\neq 0$ are outside of the contour.
To guarantee that there is a choice of such contour for which all of the inifite set of the poles $\kappa=x_1^{-m}x_2^{-m}$ with $m\ge 0$ and $\kappa=x_1^{-m-ln}x_2^{-m},x_1^{-m}x_2^{-m-ln}$ with $m\ge 0,n\ge 1$ are outside of the contour, let us consider the parameter regime $|z|=1$ and $|\mathfrak{t}|<1$.
In this regime the right-hand side can be evaluated by collecting the residues at the poles outside the contour.
Dividing both sides by ${\cal I}^{U(\infty)\text{ ADHM-}[l]}_C[\mathbb{C}^2\times \mathbb{C}^2/\mathbb{Z}_l;N](x_1;x_2)$ \eqref{N8ANinfty}, we find
\begin{align}
&\frac{
{\cal I}^{U(N)\text{ ADHM-}[l]}_C[\mathbb{C}^2\times \mathbb{C}^2/\mathbb{Z}_l;N](x_1;x_2)
}{
{\cal I}^{U(\infty)\text{ ADHM-}[l]}_C[\mathbb{C}^2\times \mathbb{C}^2/\mathbb{Z}_l;\infty](x_1;x_2)
}\nonumber \\
&=1+\sum_{m=0}^\infty (x_1x_2)^{mN}F_{m}^{(1)}
+\sum_{m=0}^\infty\sum_{n=1}^\infty \Bigl(
(x_1^{m+nl}x_2^m)^NF_{m,n}^{(2)}
+(x_1^{m}x_2^{m+nl})^NF_{m,n}^{(3)}
\Bigr),
\label{N8Asmalltnonpertexpansion}
\end{align}
with
\begin{align}
&F^{(1)}_{m}=
\biggl(
\prod_{m'=1}^m\frac{1}{1-(x_1x_2)^{-m'}}
\biggr)
\biggl(
\prod_{m'=1}^m\prod_{n'=1}^\infty
\frac{1}{1-(x_1x_2)^{-m'}x_1^{ln'}}
\frac{1}{1-(x_1x_2)^{-m'}x_2^{ln'}}
\biggr),\\
&F_{m,n}^{(2)}=
\prod_{a=0}^{m+nl-1}\frac{1}{1-(x_1x_2)^{a-m}x_1^{-nl}}
\prod_{a=0}^{nl-1}\frac{1}{1-(x_1x_2)^a(x_1^{-1}x_2)^{nl}}
\prod_{b=1}^{n-1}\prod_{a=1}^{bl-1}\frac{1}{1-(x_1x_2)^{a-bl}x_2^{bl}}\nonumber \\
&\quad\times \prod_{a=1}^{nl-1} \prod_{b=n+1}^{2n-1} \frac{1}{1-(x_1x_2)^{a-nl}x_2^{bl}}
\prod_{b=1}^{n-1} \frac{1}{1-x_1^{(b-n)l}}
\frac{1}{1-x_1^{-nl}x_2^{bl}}\nonumber \\
&\quad\times \prod_{a=0}^{m-1}\prod_{b=1}^\infty\frac{1}{1-(x_1x_2)^{a-m}x_1^{(b-n)l}}\frac{1}{1-(x_1x_2)^{a-m}x_1^{-nl}x_2^{bl}}\nonumber \\
&\quad \times \prod_{a=-nl+1}^{-1}\prod_{b=1+2n}^\infty\frac{1}{1-(x_1x_2)^ax_2^{bl}}
\prod_{b=1}^\infty\frac{1}{1-x_2^{bl}(x_1^{-1}x_2)^{nl}},\\
&F_{m,n}^{(3)}=
\prod_{a=0}^{m+nl-1}\frac{1}{1-(x_1x_2)^{a-m}x_2^{-nl}}
\prod_{a=0}^{nl-1}\frac{1}{1-(x_1x_2)^a(x_2^{-1}x_1)^{nl}}
\prod_{b=1}^{n-1}\prod_{a=1}^{bl-1}\frac{1}{1-(x_1x_2)^{a-bl}x_1^{bl}}\nonumber \\
&\quad\times \prod_{a=1}^{nl-1} \prod_{b=n+1}^{2n-1} \frac{1}{1-(x_1x_2)^{a-nl}x_1^{bl}}
\prod_{b=1}^{n-1} \frac{1}{1-x_2^{(b-n)l}}
\frac{1}{1-x_2^{-nl}x_1^{bl}}\nonumber \\
&\quad\times \prod_{a=0}^{m-1}\prod_{b=1}^\infty\frac{1}{1-(x_1x_2)^{a-m}x_2^{(b-n)l}}\frac{1}{1-(x_1x_2)^{a-m}x_2^{-nl}x_1^{bl}}\nonumber \\
&\quad \times \prod_{a=-nl+1}^{-1}\prod_{b=1+2n}^\infty\frac{1}{1-(x_1x_2)^ax_1^{bl}}
\prod_{b=1}^\infty\frac{1}{1-x_1^{bl}(x_2^{-1}x_1)^{nl}}.
\end{align}

For the single sum giant graviton expansion with respect to $x_1$, 
terms which contain an infinite product of the form
\begin{align}
\prod_{n=1}^\infty \frac{1}{1-a_n(x_1)x_2^{-b_n}},
\end{align}
with $a_n(x_1)$ any function of $x_1$ and $b_n$ positive integers, should not contribute since it is proportional to $x_2^\infty$.
See \cite{Imamura:2022aua} for a similar discussion. 

Among the above expansion coefficients $F^{(1)}_m,F^{(2)}_{m,n},F^{(3)}_{m,n}$ we find only $F^{(2)}_{0,n}$ does not suffer from such factor.
Hence we can conclude that the single sum giant graviton expansion 
is given by the right-hand side of \eqref{N8Asmalltnonpertexpansion} with only $F^{(2)}_{0,n}$ kept. 
Namely, we have 
\begin{align}
\frac{
{\cal I}^{U(N)\text{ ADHM-}[l]}_C[\mathbb{C}^2\times \mathbb{C}^2/\mathbb{Z}_l;N](x_1;x_2)
}{
{\cal I}^{U(\infty)\text{ ADHM-}[l]}_C[\mathbb{C}^2\times \mathbb{C}^2/\mathbb{Z}_l;\infty](x_1;x_2)
}
=1
+
\sum_{n=1}^\infty x_1^{nlN}F_{0,n}^{(2)}
\end{align}
with
\begin{align}
F_{0,n}^{(2)}=
\prod_{k=1}^{m}\prod_{i=0}^{l-1}\prod_{n=0}^{\infty}\frac{1}{(1-x_1^{-lk+i}x_2^{ln+i})},
\end{align}
which precisely reproduces the single sum giant graviton expansion \eqref{UNlCsinglesumgg}.

\subsection{Higgs indices}
We can repeat the same analysis for the Higgs
indices of ${\cal N}=8$ SYM theories considered in section \ref{sec_Higgs}.
For the Higgs index of ${\cal N}=8$ $U(N)$ SYM theory, the grand canonical index is given by \cite{Hayashi:2024jof}
\begin{align}
\Xi^{{\cal N}=8\, U(N)}_H(\kappa;x_1;x_2)=
\frac{1}{
(\kappa x_1;x_1)_\infty
(\kappa;x_2)_\infty
}.
\end{align}
By collecting the residues at $\kappa=x_1^{-m}$ with $m\ge 1$ and $\kappa=x_2^{-m}$ with $m\ge 0$ in the inverse transformation similar to \eqref{inversetrsf}, we find
\begin{align}
\frac{
{\cal I}^{{\cal N}=8\,U(N)}_H(x_1;x_2)
}{
{\cal I}^{{\cal N}=8\,U(\infty)}_H(x_1;x_2)
}
=1+
\sum_{m=1}^\infty x_1^{mN}F^{(1)}_m
+\sum_{m=1}^\infty x_2^{mN}F^{(2)}_m,
\end{align}
with
\begin{align}
F^{(1)}_m=\frac{
\prod_{n=1}^\infty(1-x_2^n)
}{
\prod_{n=1}^{m-1}(1-x_1^{-n})
\prod_{n=1}^{\infty}(1-x_1^{-m}x_2^n)
},\quad
F^{(2)}_m=\frac{
\prod_{n=1}^\infty(1-x_1^n)
}{
\prod_{n=1}^{m-1}(1-x_2^{-n})
\prod_{n=1}^{\infty}(1-x_2^{-m}x_1^n)
}.
\end{align}
When we expand the right-hand side in $x_2$ first, only $F^{(1)}_m$ survives, which reproduces the single sum giant graviton expansion \eqref{N8Asinglesumgg}.

For the Higgs index of ${\cal N}=8$ $USp(2N)$ SYM theory, the grand canonical index is given by \eqref{gcN8BCD}. 
By collecting the residues at $\kappa=x_1^{-2n}x_2^{-2m}$ with $m,n\ge 0$ and $\kappa=x_1^{-2n-1}x_2^{-2m-1}$ with $m,n\ge 0$ in the inverse transformation, we find
\begin{align}
\frac{
{\cal I}^{{\cal N}=8\,BCD_N}_H(x_1;x_2)
}{
{\cal I}^{{\cal N}=8\,BCD_\infty}_H(x_1;x_2)
}
=1
+\sum_{\substack{b,c\ge 0\\ (b,c)\neq (0,0)}}(x_1^{2b}x_2^{2c})^NF^{(1)}_{b,c}
+\sum_{b,c\ge 0}(x_1^{2b+1}x_2^{2c+1})^NF^{(2)}_{b,c}
\end{align}
with
${\cal I}^{{\cal N}=8\,BCD_\infty}_H(x_1;x_2)$ given as \eqref{largeNIBCD} and
\begin{align}
&F^{(1)}_{b,c}=
\prod_{b'=1}^b\prod_{c'=0}^\infty\frac{(1-x_1^{-2b'+2}x_2^{2c'+1})(1-x_1^{-2b'+1}x_2^{2c'+2})}{(1-x_1^{-2b'}x_2^{2c'})(1-x_1^{-2b'+1}x_2^{2c'+1})}\nonumber \\
&\quad\times \prod_{b'=0}^\infty\prod_{c'=1}^c\frac{(1-x_1^{2b'+2}x_2^{-2c'+1})(1-x_1^{2b'+1}x_2^{-2c'+2})}{(1-x_1^{2b'}x_2^{-2c'})(1-x_1^{2b'+1}x_2^{-2c'+1})}\nonumber \\
&\quad\times \prod_{b'=1}^b\prod_{c'=1}^c\frac{(1-x_1^{-2b'+2}x_2^{-2c'+1})(1-x_1^{-2b'+1}x_2^{-2c'+2})}{(1-x_1^{-2b'}x_2^{-2c'})(1-x_1^{-2b'+1}x_2^{-2c'+1})},
\label{calN8USp_fbc}
\\
&F^{(2)}_{b,c}=
\prod_{b'=0}^\infty\frac{1-x_1^{2b'+1}}{1-x_1^{2b'+1}x_2^{-1}}
\prod_{b'=0}^\infty\prod_{c'=1}^c\frac{(1-x_1^{2b'+1}x_2^{-2c'})(1-x_1^{2b'+2}x_2^{-2c'+1})}{(1-x_1^{2b'}x_2^{-2c'})(1-x_1^{2b'+1}x_2^{-2c'-1})}\nonumber \\
&\quad\times \prod_{c'=0}^\infty \frac{1-x_2^{2c'+1}}{1-x_1^{-1}x_2^{2c'+1}}
\prod_{b'=1}^b\prod_{c'=0}^\infty\frac{(1-x_1^{-2b'+1}x_2^{2c'+2})(1-x_1^{-2b'}x_2^{2c'+1})}{(1-x_1^{-2b'}x_2^{2c'})(1-x_1^{-2b'-1}x_2^{2c'+1})}\nonumber \\
&\quad\times \frac{1}{1-x_1^{-1}x_2^{-1}}
\prod_{b'=1}^b\frac{1-x_1^{-2b'+1}}{1-x_1^{-2b'-1}x_2^{-1}}
\prod_{c'=1}^b\frac{1-x_1^{-2c'+1}}{1-x_1^{-1}x_2^{-2c'-1}}\nonumber \\
&\quad\times \prod_{b'=1}^b\prod_{c'=1}^c\frac{(1-x_1^{-2b'+1}x_2^{-2c'})(1-x_1^{-2b'}x_2^{-2c'+1})}{(1-x_1^{-2b'}x_2^{-2c'})(1-x_1^{-2b'-1}x_2^{-2c'-1})}.\label{calN8USp_fbc2}
\end{align}
When we expand the right-hand side with respect to $x_2$ first, only $F^{(1)}_{b,0}$ survives, which reproduces the single sum giant graviton expansion \eqref{N8BCDsinglesumgg}.


For the Higgs index of the $\mathcal{N}=8$ $SO(2N)$ SYM theory, the grand canonical index is given as \eqref{gcN8SO}.
By collecting the residues at $\kappa=x_1^{-2b}x_2^{-2c}, x_1^{-2b-1}x_2^{-2c-1}, x_1^{-2b}x_2^{-2c-1}, x_1^{-2b-1}x_2^{-2c}$ ($b,c\ge 0$),
we find
\be\begin{split}
&\frac{\mathcal{I}_H^{\mathcal{N}=8\;SO(2N)}(x_1; x_2) }{\mathcal{I}_H^{\mathcal{N}=8\;SO(\infty)}(x_1; x_2)}\cr
&\hspace{1cm}= 1 + \sum_{\substack{b, c\geq 0\\ (b, c) \neq (0,0)}}\left(x_1^{2b}x_2^{2c}\right)^NF_{b, c}^{(1)}(x_1; x_2)+ \sum_{b, c\geq 0}\left(x_1^{2b+1}x_2^{2c+1}\right)^NF_{b, c}^{(2)}(x_1; x_2)\cr
&\hspace{2cm}  + \sum_{b, c\geq 0}\left(x_1^{2b}x_2^{2c+1}\right)^NF_{b, c}^{(3)}(x_1; x_2) + \sum_{b, c\geq 0}\left(x_1^{2b+1}x_2^{2c}\right)^NF_{c, b}^{(3)}(x_2; x_1),
\end{split}
\label{inversetrsfSO}
\ee
where
\begin{align}
&F_{b, c}^{(1)}(x_1; x_2)
 =\prod_{b'=1}^b\prod_{c'=0}^{\infty}\frac{\left(1-x_1^{-2b'+2}x_2^{2c'+1}\right)\left(1-x_1^{-2b'+1}x_2^{2c'+2}\right)}{\left(1-x_1^{-2b'}x_2^{2c'}\right)\left(1-x_1^{-2b'+1}x_2^{2c'+1}\right)}\nonumber \\
&\quad\times \prod_{b'=0}^{\infty}\prod_{c'=1}^c\frac{\left(1-x_1^{2b'+2}x_2^{-2c'+1}\right)\left(1-x_1^{2b'+1}x_2^{-2c'+2}\right)}{\left(1-x_1^{2b'}x_2^{-2c'}\right)\left(1-x_1^{2b'+1}x_2^{-2c'+1}\right)}\cr
&\quad\times \prod_{b'=1}^b\prod_{c'=1}^c\frac{\left(1-x_1^{-2b'+2}x_2^{-2c'+1}\right)\left(1-x_1^{-2b'+1}x_2^{-2c'+2}\right)}{\left(1-x_1^{-2b'}x_2^{-2c'}\right)\left(1-x_1^{-2b'+1}x_2^{-2c'+1}\right)},\\
&F_{b, c}^{(2)}(x_1; x_2)
=\prod_{b'=1}^b\prod_{c'=0}^{\infty}\frac{\left(1-x_1^{-2b'}x_2^{2c'+1}\right)\left(1-x_1^{-2b'+1}x_2^{2c'}\right)}{\left(1-x_1^{-2b'}x_2^{2c'}\right)\left(1-x_1^{-2b'-1}x_2^{2c'+1}\right)}\nonumber \\
&\quad\times \prod_{b'=0}^{\infty}\prod_{c'=1}^c\frac{\left(1-x_1^{2b'}x_2^{-2c'+1}\right)\left(1-x_1^{2b'+1}x_2^{-2c'}\right)}{\left(1-x_1^{2b'}x_2^{-2c'}\right)\left(1-x_1^{2b'+1}x_2^{-2c'-1}\right)}\cr
&\quad\times \prod_{b'=1}^b\prod_{c'=1}^c\frac{\left(1-x_1^{-2b'}x_2^{-2c'+1}\right)\left(1-x_1^{-2b'+1}x_2^{-2c'}\right)}{\left(1-x_1^{-2b'}x_2^{-2c'}\right)\left(1-x_1^{-2b'-1}x_2^{-2c'-1}\right)}\prod_{c'=0}^{\infty}\frac{1-x_2^{2c'+1}}{1-x_1^{-1}x_2^{2c'+1}}\prod_{b'=0}^{\infty}\frac{1-x_1^{2b'+1}}{1-x_1^{2b'+1}x_2^{-1}}\cr
&\quad\times \frac{1}{1-x_1^{-1}x_2^{-1}}\prod_{b'=1}^b\frac{1}{1-x_1^{-2b'-1}x_2^{-1}}\prod_{c'=1}^c\frac{1}{1-x_1^{-1}x_2^{-2c'-1}},
\end{align}
and
\begin{align}
&F_{b, c}^{(3)}(x_1; x_2)
=\prod_{b'=1}^b\prod_{c'=0}^{\infty}\frac{\left(1-x_1^{-2b'+1}x_2^{2c'}\right)\left(1-x_1^{-2b'+2}x_2^{2c'+1}\right)}{\left(1-x_1^{-2b'}x_2^{2c'}\right)\left(1-x_1^{-2b'+1}x_2^{2c'+1}\right)}\nonumber \\
&\quad\times \prod_{b'=0}^{\infty}\prod_{c'=1}^c\frac{\left(1-x_1^{2b'+1}x_2^{-2c'}\right)\left(1-x_1^{2b'+2}x_2^{-2c'+1}\right)}{\left(1-x_1^{2b'}x_2^{-2c'}\right)\left(1-x_1^{2b'+1}x_2^{-2c'-1}\right)}\cr
&\quad\times \prod_{b'=1}^b\prod_{c'=1}^c\frac{\left(1-x_1^{-2b'+1}x_2^{-2c'}\right)\left(1-x_1^{-2b'+2}x_2^{-2c'+1}\right)}{\left(1-x_1^{-2b'}x_2^{-2c'}\right)\left(1-x_1^{-2b'+1}x_2^{-2c'-1}\right)}\prod_{b'=0}^{\infty}\frac{1-x_1^{2b'+1}}{1-x_1^{2b'+1}x_2^{-1}}\nonumber \\
&\quad\times \prod_{b'=1}^b\frac{1}{1-x_1^{-2b'+1}x_2^{-1}}.
\end{align}
$F^{(1)}_{b,c}(x_1;x_2)$ and $F^{(2)}_{b,c}(x_1;x_2)$ are the same as \eqref{calN8USp_fbc} and \eqref{calN8USp_fbc2} respectively. When we expand the right-hand side of \eqref{inversetrsfSO} in terms of $x_2$ and truncate the series at some order, only the terms with $F_{b,0}^{(1)}(x_1;x_2)$ and $F_{0,b}^{(3)}(x_2;x_1)$ appear at the finite order of $x_2$.
Hence we obtain the single sum expansion as \eqref{SO2Nsinglesumgg}.

\section{Nekrasov formulas for Higgs indices of $U(N)$ ADHM theories}
\label{app_UNlHgg}
In this appendix we explain how we obtain the leading terms in the giant graviton expansions of the Higgs indices of $U(N)$ ADHM theory with $l$ flavors \eqref{GGind_ADHM_H}.
First, the Higgs indices of these theories can be calculated by the JK residue sum \cite{MR1318878}, which result in the Nekrasov formula for the instanton partition function of 5d ${\cal N}=1$ Yang-Mills theory with gauge group $SU(l)$ and the instanton number $N$ \cite{Nekrasov:2002qd,Nekrasov:2003rj,Hwang:2014uwa}, up to an overall factor, as
\begin{align}
{\cal I}^{U(N)\text{ADHM-}[l]}_H[\mathbb{C}^2\times \mathbb{C}^2/\mathbb{Z}_l;N](x_1;x_2)=\frac{1}{(x_1x_2)^{Nl}}\sum_{
\substack{
\lambda^{(1)},\cdots,\lambda^{(l)}\\
\sum_{\alpha=1}^l|\lambda^{(\alpha)}|=N
}}
\prod_{\alpha,\beta=1}^l\frac{1}{{\cal N}_{\lambda^{(\alpha)},\lambda^{(\beta)}}(\frac{y^{(\beta)}}{y^{(\alpha)}})},
\label{Nekrasov}
\end{align}
where $y_\alpha$ are the fugacities coupled to the flavor charges and
\begin{align}
{\cal N}_{\lambda,\mu}(u)=
\prod_{\Box\in\lambda}(1-u
x_1^{-\text{leg}_\mu(\Box)-1}
x_2^{\text{arm}_\lambda(\Box)}
)
\prod_{\Box'\in\mu}(1-u
x_1^{\text{leg}_\lambda(\Box')}
x_2^{-\text{arm}_\mu(\Box')-1}
).
\end{align}
Using this formula we can generate ${\cal I}^{U(N)\text{ADHM-}[l]}_H(x_1=x^{-1}\mathfrak{t};x_2=x\mathfrak{t})$ in small $\mathfrak{t}$ expansion for various $l$ and $N$.
Hereafter we set the fugacities $y_\alpha$ for the flavor symmetry to $y_\alpha=1$.
Dividing the Higgs indices by the Higgs indices at $N=\infty$ \cite{Crew:2020psc}
\begin{align}
&{\cal I}^{U(\infty)\text{ADHM-}[l]}_H[\mathbb{C}^2\times \mathbb{C}^2/\mathbb{Z}_l;N=\infty](x_1=x^{-1}\mathfrak{t};x_2=x\mathfrak{t})\nonumber \\
&=\prod_{n=0}^\infty\prod_\pm \frac{1}{1-(x^{\pm 1}\mathfrak{t})^{n+1}}
\prod_{m,n=1}^\infty\frac{1}{(1-x^{m-n}\mathfrak{t}^{m+n})^{l^2}}
\end{align}
and expanding it again in $\mathfrak{t}$, we obtain
{\fontsize{9pt}{1pt}\selectfont
\begin{align}
&\frac{{\cal I}^{U(N)\text{ADHM-}[2]}_H[\mathbb{C}^2\times \mathbb{C}^2/\mathbb{Z}_l;N](x_1=x^{-1}\mathfrak{t};x_2=x\mathfrak{t})}
{{\cal I}^{U(\infty)\text{ADHM-}[2]}_H[\mathbb{C}^2\times \mathbb{C}^2/\mathbb{Z}_l;N=\infty](x_1=x^{-1}\mathfrak{t};x_2=x\mathfrak{t})}
=1+\mathfrak{t}^N
\biggl[
\Bigl(-\frac{x^{N+2}-x^{-N-2}}{x-x^{-1}}\Bigr)\mathfrak{t}\nonumber \\
&\quad \quad +\Bigl(-\frac{(x^2+3+x^{-2})(x^{N+1}-x^{-N-1})}{x-x^{-1}}\Bigr)\mathfrak{t}^2\nonumber \\
&\quad\quad +\Bigl(-\frac{x^N(x^4+3x^2+5+4x^{-2}+x^{-4})-(x\rightarrow x^{-1})}{x-x^{-1}}\Bigr)\mathfrak{t}^3\nonumber \\
&\quad\quad +\Bigl(-\frac{x^N(x^5+3x^3+5x+11x^{-1}+9x^{-3}+4x^{-5}+x^{-7})-(x\rightarrow x^{-1})}{x-x^{-1}}\Bigr)\mathfrak{t}^4\nonumber \\
&\quad\quad +\Bigl(-\frac{
x^N (x^6 + 3 x^4 + 5 x^2 + 11 + 22 x^{-2} + 19 x^{-4} + 10 x^{-6} + 4 x^{-8} + x^{-10})
-(x\rightarrow x^{-1})
}{x-x^{-1}}
\Bigr)\mathfrak{t}^5\nonumber \\
&\quad\quad
+\Bigl[-\frac{1}{x-x^{-1}}
\Bigl(
x^N (x^7 + 3 x^5 + 5 x^3 + 11 x + 22 x^{-1} + 38 x^{-3} + 38 x^{-5} + 23 x^{-7} + 10 x^{-9} + 4 x^{-11} + x^{-13})\nonumber \\
&\quad\quad\quad\quad\quad\quad\quad\quad -(x\rightarrow x^{-1})
\Bigr)
\Bigr]\mathfrak{t}^6
\nonumber \\
&\quad\quad
+\Bigl[-\frac{1}{x-x^{-1}}
\Bigl(
x^N (
x^8
+ 3 x^6 
+ 5 x^4 
+ 11 x^2 
+ 22 
+ 38x^{-2}
+ 67x^{-4}
+ 70x^{-6}
+ 47x^{-8}
+ 24x^{-10}
+ 10x^{-12}\nonumber \\
&\quad\quad\quad\quad\quad\quad\quad\quad + 4x^{-14}
+ x^{-16}
)
-(x\rightarrow x^{-1})
\Bigr)\Bigr]\mathfrak{t}^7
\nonumber \\
&\quad\quad
+\Bigl[-\frac{1}{x-x^{-1}}
\Bigl(
x^N (
x^9
+ 3 x^7
+ 5 x^5
+ 11 x^3
+ 22 x
+ 38x^{-1}
+ 67x^{-3}
+ 113x^{-5}
+ 123x^{-7}
+ 90x^{-9}
+ 51x^{-11}
\nonumber \\
&\quad\quad\quad\quad\quad\quad\quad\quad
+ 24x^{-13}
+ 10x^{-15}
+ 4x^{-17}
+ x^{-19}
)
-(x\rightarrow x^{-1})
\Bigr)\Bigr]\mathfrak{t}^8
\nonumber \\
&\quad\quad
+\Bigl[-\frac{1}{x-x^{-1}}
\Bigl(
x^N (
x^{10}
+ 3 x^8 
+ 5 x^6 
+ 11 x^4 
+ 22 x^2 
+ 38 
+ 67x^{-2}
+ 113x^{-4}
+ 185x^{-6}
+ 209x^{-8}
+ 164x^{-10}\nonumber \\
&\quad\quad\quad\quad\quad\quad\quad\quad
+ 99x^{-12}
+ 52x^{-14}
+ 24x^{-16}
+ 10x^{-18}
+ 4x^{-20}
+ x^{-22}
)
-(x\rightarrow x^{-1})
\Bigr)\Bigr]\mathfrak{t}^9
\nonumber \\
&\quad\quad
+\Bigl[-\frac{1}{x-x^{-1}}
\Bigl(
x^N (
x^{11}
+ 3 x^9 
+ 5 x^7 
+ 11 x^5 
+ 22 x^3 
+ 38 x 
+ 67x^{-1}
+ 113x^{-3}
+ 185x^{-5}
+ 299x^{-7}
+ 345x^{-9}\nonumber \\
&\quad\quad\quad\quad\quad\quad\quad\quad
+ 284x^{-11}
+ 185x^{-13}
+ 103x^{-15}
+ 52x^{-17}
+ 24x^{-19}
+ 10x^{-21}
+ 4x^{-23}
+ x^{-25}
)\nonumber \\
&\quad\quad\quad\quad\quad\quad\quad\quad -(x\rightarrow x^{-1})
\Bigr)\Bigr]\mathfrak{t}^{10}
\nonumber \\
&\quad\quad
+\Bigl[-\frac{1}{x-x^{-1}}
\Bigl(
x^N (
x^{12}
+ 3 x^{10}
+ 5 x^8 
+ 11 x^6 
+ 22 x^4 
+ 38 x^2 
+ 67 
+ 113x^{-2}
+ 185x^{-4}
+ 299x^{-6}
+ 471x^{-8}\nonumber \\
&\quad\quad\quad\quad\quad\quad\quad\quad + 555x^{-10}
+ 479x^{-12}
+ 329x^{-14}
+ 194x^{-16}
+ 104x^{-18}
+ 52x^{-20}
+ 24x^{-22}
+ 10x^{-24}
+ 4x^{-26}\nonumber \\
&\quad\quad\quad\quad\quad\quad\quad\quad + x^{-28}
)
-(x\rightarrow x^{-1})
\Bigr)\Bigr]\mathfrak{t}^{11}
\nonumber \\
&\quad\quad +{\cal O}(\mathfrak{t}^{12})
\biggr]+{\cal O}(\mathfrak{t}^{2N+4}),\\
%
%
&\frac{{\cal I}^{U(N)\text{ADHM-}[3]}_H[\mathbb{C}^2\times \mathbb{C}^2/\mathbb{Z}_l;N](x_1=x^{-1}\mathfrak{t};x_2=x\mathfrak{t})}
{{\cal I}^{U(\infty)\text{ADHM-}[3]}_H[\mathbb{C}^2\times \mathbb{C}^2/\mathbb{Z}_l;N=\infty](x_1=x^{-1}\mathfrak{t};x_2=x\mathfrak{t})}
=1+\mathfrak{t}^N
\biggl[
\Bigl(-\frac{x^{N+2}-x^{-N-2}}{x-x^{-1}}\Bigr)\mathfrak{t}\nonumber \\
&\quad\quad +\Bigl(-\frac{
x^N (
x^3
+ 8 x
+ x^{-1}
)
-(x\rightarrow x^{-1})
}{x-x^{-1}}
\Bigr)\mathfrak{t}^2
\nonumber \\
&\quad\quad +\Bigl(-\frac{
x^N (
 x^4
 + 8 x^2
+ 18
 + 9x^{-2}
 + x^{-4}
)
-(x\rightarrow x^{-1})
}{x-x^{-1}}
\Bigr)\mathfrak{t}^3\nonumber \\
&\quad\quad +\Bigl(-\frac{
x^N (
x^5
+ 8 x^3 
+ 18 x 
+ 55x^{-1}
+ 27x^{-3}
+ 9x^{-5}
+ x^{-7}
)
)
-(x\rightarrow x^{-1})
}{x-x^{-1}}
\Bigr)\mathfrak{t}^4\nonumber \\
&\quad\quad +\Bigl(-\frac{
x^N (
x^6
+ 8 x^4 
+ 18 x^2 
+ 55 
+ 125x^{-2}
+ 81x^{-4}
+ 28x^{-6}
+ 9x^{-8}
+ x^{-10}
)
)
-(x\rightarrow x^{-1})
}{x-x^{-1}}
\Bigr)\mathfrak{t}^5\nonumber \\
&\quad\quad +\Bigl[-\frac{
1
}{x-x^{-1}}
\Bigl(x^N (
x^7
+ 8 x^5 
+ 18 x^3 
+ 55 x 
+ 125x^{-1}
+ 279x^{-3}
+ 198x^{-5}
+ 90x^{-7}
+ 28x^{-9}
+ 9x^{-11}\nonumber \\
&\quad\quad\quad\quad\quad\quad\quad\quad + x^{-13}
)
-(x\rightarrow x^{-1})
\Bigr)
\Bigr]\mathfrak{t}^6\nonumber \\
&\quad\quad +\Bigl[-\frac{
1
}{x-x^{-1}}
\Bigl(
x^N (
x^8 
+ 8 x^6 
+ 18 x^4 
+ 55 x^2 
+ 125 
+ 279x^{-2}
+ 569x^{-4}
+ 458x^{-6}
+ 225x^{-8}
+ 91x^{-10}\nonumber \\
&\quad\quad\quad\quad\quad\quad\quad\quad + 28x^{-12}
+ 9x^{-14}
+ x^{-16}
)
-(x\rightarrow x^{-1})
\Bigr)
\Bigr]\mathfrak{t}^7
+{\cal O}(\mathfrak{t}^{8})
\biggr]+{\cal O}(\mathfrak{t}^{2N+4}),\\
%
%
&\frac{{\cal I}^{U(N)\text{ADHM-}[4]}_H[\mathbb{C}^2\times \mathbb{C}^2/\mathbb{Z}_l;N](x_1=x^{-1}\mathfrak{t};x_2=x\mathfrak{t})}
{{\cal I}^{U(\infty)\text{ADHM-}[4]}_H[\mathbb{C}^2\times \mathbb{C}^2/\mathbb{Z}_l;N=\infty](x_1=x^{-1}\mathfrak{t};x_2=x\mathfrak{t})}
=1+\mathfrak{t}^N
\biggl[
\Bigl(-\frac{x^{N+2}-x^{-N-2}}{x-x^{-1}}\Bigr)\mathfrak{t}\nonumber \\
&\quad\quad+\Bigl(-\frac{
x^N (
x^3
+ 15 x
+ x^{-1}
)
-(x\rightarrow x^{-1})
}{x-x^{-1}}
\Bigr)\mathfrak{t}^2
\nonumber \\
&\quad\quad +\Bigl(-\frac{
x^N (
 x^4
 + 15 x^2
+ 52
 + 16x^{-2}
 + x^{-4}
)
-(x\rightarrow x^{-1})
}{x-x^{-1}}
\Bigr)\mathfrak{t}^3\nonumber \\
&\quad\quad +\Bigl(-\frac{
x^N (
x^5
+ 15 x^3 
+ 52 x 
+ 188x^{-1}
+ 68x^{-3}
+ 16x^{-5}
+ x^{-7}
)
)
-(x\rightarrow x^{-1})
}{x-x^{-1}}
\Bigr)\mathfrak{t}^4\nonumber \\
&\quad\quad +\Bigl(-\frac{
x^N (
x^6
+ 15 x^4 
+ 52 x^2 
+ 188
+ 521x^{-2}
+ 255x^{-4}
+ 69x^{-6}
+ 16x^{-8}
+ x^{-10}
)
)
-(x\rightarrow x^{-1})
}{x-x^{-1}}
\Bigr)\mathfrak{t}^5\nonumber \\
&\quad\quad +\Bigl[-\frac{
1
}{x-x^{-1}}
\Bigl(x^N (
x^7
+ 15 x^5 
+ 52 x^3 
+ 188 x 
+ 521x^{-1}
+ 1383x^{-3}
+ 761x^{-5}
+ 271x^{-7}
+ 69x^{-9}\nonumber \\
&\quad\quad\quad\quad\quad\quad\quad\quad + 16x^{-11}
+ x^{-13}
)
-(x\rightarrow x^{-1})
\Bigr)
\Bigr]\mathfrak{t}^6
+{\cal O}(\mathfrak{t}^{7})
\biggr]+{\cal O}(\mathfrak{t}^{2N+4}),\\
%
%
&\frac{{\cal I}^{U(N)\text{ADHM-}[5]}_H[\mathbb{C}^2\times \mathbb{C}^2/\mathbb{Z}_l;N](x_1=x^{-1}\mathfrak{t};x_2=x\mathfrak{t})}
{{\cal I}^{U(\infty)\text{ADHM-}[5]}_H[\mathbb{C}^2\times \mathbb{C}^2/\mathbb{Z}_l;N=\infty](x_1=x^{-1}\mathfrak{t};x_2=x\mathfrak{t})}
=1+\mathfrak{t}^N
\biggl[
\Bigl(-\frac{x^{N+2}-x^{-N-2}}{x-x^{-1}}\Bigr)\mathfrak{t}\nonumber \\
&\quad\quad +\Bigl(-\frac{
x^N (
x^3
+ 24 x
+ x^{-1}
)
-(x\rightarrow x^{-1})
}{x-x^{-1}}
\Bigr)\mathfrak{t}^2
\nonumber \\
&\quad\quad +\Bigl(-\frac{
x^N (
 x^4
 + 24 x^2
+ 125
 + 25x^{-2}
 + x^{-4}
)
-(x\rightarrow x^{-1})
}{x-x^{-1}}
\Bigr)\mathfrak{t}^3\nonumber \\
&\quad\quad +\Bigl(-\frac{
x^N (
x^5
+ 24 x^3 
+ 125 x 
+ 525x^{-1}
+ 150x^{-3}
+ 25x^{-5}
+ x^{-7}
)
)
-(x\rightarrow x^{-1})
}{x-x^{-1}}
\Bigr)\mathfrak{t}^4\nonumber \\
&\quad\quad +\Bigl[-\frac{
1}{x-x^{-1}}
\Bigl(
x^N \Bigl(
x^6
+ 24 x^4 
+ 125 x^2 
+ 525
+ 1775x^{-2}
+ 674x^{-4}
+ 151x^{-6}
+ 25x^{-8}\nonumber \\
&\quad\quad\quad\quad\quad\quad\quad\quad + x^{-10}
)
\Bigr)
-(x\rightarrow x^{-1})
\Bigr]\mathfrak{t}^5
+{\cal O}(\mathfrak{t}^6)
\biggr]+{\cal O}(\mathfrak{t}^{2N+4}).
\end{align}
}
From these results we observe that the coefficients of $\mathfrak{t}^N$ take the following form for any $l$ (including $l=0,1$):
\begin{align}
&\frac{
{\cal I}^{U(N)\text{ADHM-}[l]}_H[\mathbb{C}^2\times \mathbb{C}^2/\mathbb{Z}_l;N](x_1;x_2)
}{
{\cal I}^{U(\infty)\text{ADHM-}[l]}_H[\mathbb{C}^2\times \mathbb{C}^2/\mathbb{Z}_l;N=\infty](x_1;x_2)
}=1+
x_1^{N}{\hat {\cal F}}_1^{(l)}(x_1;x_2)
+x_2^{N}{\hat {\cal F}}_1^{(l)}(x_2;x_1)\nonumber \\
&\quad +{\cal O}(\mathfrak{t}^{2N}),
\label{UNlleadingggexp}
\end{align}
where
\begin{align}
{\hat {\cal F}}^{(l)}_1(x_1;x_2)=\frac{
h^{(l)}(x_2)
}{(x_1^{-1};x_2)_\infty}
\label{assumedstructureofcalFl1}
\end{align}
with $h^{(l)}(x_2)$ some function depending only on $x_2=x\mathfrak{t}$.

Once we assume the structure \eqref{assumedstructureofcalFl1}, we can read off $h^{(l)}(x_2)$ even from the expansion coefficients of the Higgs indices in the unrefined limit ${\cal I}^{U(N)\text{ADHM-}[l]}_H(x_1=x_2=\mathfrak{t})$, which can be obtained realtively easily by evaluating the Nekrasov formula \eqref{Nekrasov} numerically at a high precision.
Suppose that the coefficient of $\mathfrak{t}^N$ in the unrefined limit is
\begin{align}
\frac{
{\cal I}^{U(N)\text{ADHM-}[l]}_H[\mathbb{C}^2\times \mathbb{C}^2/\mathbb{Z}_l;N](x_1=x_2=\mathfrak{t})
}{
{\cal I}^{U(\infty)\text{ADHM-}[l]}_H[\mathbb{C}^2\times \mathbb{C}^2/\mathbb{Z}_l;N](x_1=x_2=\mathfrak{t})
}=1+\sum_{n=1}^\infty(a_nN+b_n)\mathfrak{t}^{n+N}+{\cal O}(\mathfrak{t}^{2N})
\end{align}
with some coefficients $\{a_n,b_n\}$.
Then, from \eqref{UNlleadingggexp} and \eqref{assumedstructureofcalFl1} it follows that
\begin{align}
&\sum_{n=1}^\infty(a_nN+b_n)\mathfrak{t}^n\nonumber \\
&=\lim_{x\rightarrow 1}(
x^{-N}
{\cal F}^{(l)}_1(x^{-1}\mathfrak{t};x\mathfrak{t})
+
x^N
{\cal F}^{(l)}_1(x\mathfrak{t};x^{-1}\mathfrak{t})
)\nonumber \\
&=\biggl(\prod_{\substack{n=0\\ (n\neq 1)}}^\infty\frac{1}{1-\mathfrak{t}^{n-1}}\biggr)h(\mathfrak{t})\biggl[N+1+\sum_{\substack{n=0\\ (n\neq 1)}}^\infty \Bigl(-\frac{(n+1)\mathfrak{t}^{n-1}}{1-\mathfrak{t}^{n-1}}\Bigr)-\frac{\mathfrak{t}}{h(\mathfrak{t})}\frac{dh(x_2')}{dx_2'}\Bigr|_{x_2'=\mathfrak{t}}\biggr].
\end{align}
Hence we obtain $h(x_2)$ as
\begin{align}
h(x_2)=-(1-x_2)\prod_{n=1}^\infty(1-x_2^n)\sum_{n=1}^\infty a_nx_2^{n-1}.
\end{align}
Here we list the final results for $l\le 14$
{\fontsize{9pt}{1pt}\selectfont
\begin{align}
&h^{(l=0)}(x_2)=(x_2;x_2)_\infty,\nonumber \\
&h^{(l=1)}(x_2)=1,\nonumber \\
&h^{(l=2)}(x_2)=1 + 3 x_2 + 4 x_2^{2} + 7 x_2^{3} + 13 x_2^{4} + 19 x_2^{5} + 29 x_2^{6} + 43 x_2^{7} + 62 x_2^{8} + 90 x_2^{9} + 126 x_2^{10} + 174 x_2^{11} + 239 x_2^{12}\nonumber \\
&\quad + 325 x_2^{13} + 435 x_2^{14} + 580 x_2^{15} + 769 x_2^{16} + 1007 x_2^{17} + 1313 x_2^{18} + 1702 x_2^{19}+\cdots,\nonumber \\
&h^{(l=3)}(x_2)=1 + 8 x_2 + 17 x_2^2 + 46 x_2^3 + 98 x_2^4 + 198 x_2^5 + 371 x_2^6 + 692 x_2^7 + 1205 x_2^8 +2082 x_2^9
+3463 x_2^{10} +5678 x_2^{11}\nonumber \\
&\quad +9085 x_2^{12} + 14370 x_2^{13} + 22273 x_2^{14} + 34178 x_2^{15} +51674 x_2^{16} +77362 x_2^{17} +114452 x_2^{18} +167916 x_2^{19} + \cdots,\nonumber \\
&h^{(l=4)}(x_2)=1 + 15 x_2 + 51 x_2^2 + 172 x_2^3 + 453 x_2^4 + 1128 x_2^5 + 2539 x_2^6 + 5505 x_2^7 + 11238 x_2^8 + 22259 x_2^9 +42438 x_2^{10}\nonumber \\
&\quad + 78909 x_2^{11} + 142770 x_2^{12} +\cdots,\nonumber \\
&h^{(l=5)}(x_2)=1 + 24 x_2 + 124 x_2^2 + 500 x_2^3 + 1625 x_2^4 + 4752 x_2^5 + 12524 x_2^6 + 31000 x_2^7 +72250 x_2^8 +161000 x_2^9\nonumber \\
&\quad + 344378 x_2^{10} + 712548 x_2^{11}+ \cdots,\nonumber \\
&h^{(l=6)}(x_2)=1 + 35 x_2 + 260 x_2^2 + 1255 x_2^3 + 4910 x_2^4 + 16566 x_2^5 + 50175 x_2^6 +140140 x_2^7 +366565 x_2^8 +908165 x_2^9\nonumber \\
&\quad + \cdots,\nonumber \\
&h^{(l=7)}(x_2)=1 + 48 x_2 + 489 x_2^2 + 2842 x_2^3 + 13083 x_2^4 + 50520 x_2^5 +173362 x_2^6 + 541452 x_2^7+1571919 x_2^8 +\cdots,\nonumber \\
&h^{(l=8)}(x_2)=1 + 63 x_2 + 847 x_2^2 + 5936 x_2^3 + 31668 x_2^4+139069 x_2^5+535164 x_2^6+1855505 x_2^7+ 5931849 x_2^8 +\cdots,\nonumber \\
&h^{(l=9)}(x_2)=1 + 80 x_2 + 1376 x_2^2 + 11592 x_2^3 + 71028 x_2^4 +352504 x_2^5 +1510272 x_2^6 + 5777704 x_2^7 +20218473 x_2^8\nonumber \\
&\quad +\cdots,\nonumber \\
&h^{(l=10)}(x_2)=1 + 99 x_2 + 2124 x_2^2 + 21375 x_2^3 + 149625 x_2^4 + 834255 x_2^5+ 3958521 x_2^6 +16623747 x_2^7\nonumber \\
&\quad +63376875 x_2^8 + \cdots,\nonumber \\
&h^{(l=11)}(x_2)=1 + 120 x_2 + 3145 x_2^2 + 37510 x_2^3 + 298870 x_2^4 + 1862554 x_2^5 + 9747155 x_2^6+ 44740840 x_2^7\nonumber \\
&\quad  +185107010 x_2^8 + \cdots,\nonumber \\
&h^{(l=12)}(x_2)=1 + 143 x_2 + 4499 x_2^2 + 63052 x_2^3 + 570053 x_2^4 + 3953928 x_2^5 + 22740971 x_2^6 +113686529 x_2^7\nonumber \\
&\quad  +508760615 x_2^8 +\cdots,\nonumber \\
&h^{(l=13)}(x_2)=1 + 168 x_2 + 6252 x_2^2 + 102076 x_2^3 + 1043913 x_2^4 + 8030712 x_2^5 +50607388 x_2^6 + 274712712 x_2^7\nonumber \\
&\quad +1325979915 x_2^8 +\cdots,\nonumber \\
&h^{(l=14)}(x_2)=1 + 195 x_2 + 8476 x_2^2 + 159887 x_2^3 + 1843478 x_2^4 + 15683382 x_2^5 + 107993795 x_2^6 + 634906480 x_2^7\nonumber \\
&\quad + 3297231262 x_2^8+\cdots.
\end{align}
}
Interestingly, we observe that $h^{(l)}(x_2)$ agree with the vacuum character of the $SU(l)_{1}$ WZW model \cite{DiFrancesco:1997nk}
\begin{align}
h^{(l)}(x_2)&
=\chi_{\widehat{\mathfrak{su}}(l)_{1}}(x_2)
\nonumber\\
&=\frac{1}{(x_2;x_2)_{\infty}^{l-1}}
\sum_{
\begin{smallmatrix}
m_1,\cdots, m_l\in \mathbb{Z}\\
m_1+m_2+\cdots+m_l=0\\
\end{smallmatrix}
}x_2^{\sum_{i=1}^{l}\frac{m_i^2}{2}}. 
\end{align}
After multiplication by the generating function for the ordinary partitions, 
it becomes the generating function for the generalized Frobenius partitions $c\phi_l(n)$ of $n$ with $l$ colors \cite{MR743546}
\begin{align}
\sum_{n=0}^{\infty} c\phi_{l}(n)q^n&=
\prod_{n=1}^{\infty}\frac{1}{1-q^n}
\chi_{\widehat{\mathfrak{su}}(l)_{1}}(q)
\nonumber\\
&=\oint\frac{dz}{2\pi iz}\prod_{i=0}^\infty (1+zq^i)^l(1+z^{-1}q^{i+1})^l. 
\end{align}

\bibliographystyle{utphys}
\bibliography{ref}

\end{document}